\documentclass[journal]{IEEEtran}
\IEEEoverridecommandlockouts

\usepackage[T1]{fontenc}
\usepackage{amsmath}
\usepackage{amssymb}
\usepackage{amsfonts}
\usepackage{graphicx}
\usepackage{pifont}
\usepackage{algorithm}
\usepackage{algpseudocode}
\usepackage{url}
\usepackage{xcolor}
\usepackage{multirow}
\usepackage{array}
\usepackage{cite}

\graphicspath{{Figures/}}

\newcommand{\cmark}{\ding{51}}
\newcommand{\xmark}{\ding{55}}
\newcommand{\bs}[1]{\boldsymbol{#1}}
\newcommand{\etal}{\emph{et al.}}
\newcommand{\Comp}{\mathbb{C}}
\newcommand{\Exp}{\mathbb{E}}
\newcommand{\sig}{\operatorname{sig}}
\DeclareMathOperator{\diag}{diag}

\newtheorem{proposition}{Proposition}
\newtheorem{corollary}{Corollary}
\newtheorem{remark}{Remark}
\newtheorem{assumption}{Assumption}

\usepackage{hyperref}
\hypersetup{colorlinks=true, citecolor=blue, linkcolor=blue, urlcolor=blue}

\begin{document}
\bstctlcite{IEEEexample:BSTcontrol}

\title{Certified Covertness Without Warden CSI in STAR-RIS-Assisted RSMA Massive MIMO: Closed-Form and DRL-Based Designs}

\author{Pham~Quang~Minh, Nguyen Thai Anh, Huy T. Nguyen,~\IEEEmembership{Member,~IEEE}, Tri-Hai Nguyen,~\IEEEmembership{Member,~IEEE}, Thai-Hoc Vu, Tran Thien Thanh,~\IEEEmembership{Member,~IEEE}, Vo Nguyen Quoc Bao,~\IEEEmembership{Senior Member,~IEEE}, and Ngo~Hoang~Tu,~\IEEEmembership{Member,~IEEE}%
\thanks{This work is funded by the National Foundation for Science and Technology Development (NAFOSTED) under Grant number 102.02-2025.97. \textit{(Corresponding author: Ngo Hoang Tu.)}}
\thanks{Pham Quang Minh, Nguyen Thai Anh, Huy T. Nguyen, Tri-Hai Nguyen, Vo Nguyen Quoc Bao, and Ngo Hoang Tu are with the Faculty of Information Technology, Van Lang School of Technology, Van Lang University, Ho Chi Minh City 70000, Vietnam (e-mail: minh.2274802010538@vanlanguni.vn, anh.nt@vlu.edu.vn, huy.nt@vlu.edu.vn, hai.nguyentri@vlu.edu.vn, bao.vnq@vlu.edu.vn, tu.nh@vlu.edu.vn).}
\thanks{Thai-Hoc Vu is with the Faculty of Electrical Engineering and Computer Science, VSB-Technical University of Ostrava, 17. Listopadu 2172/15, 708 00, Ostrava, Czechia (e-mail: thai.hoc.vu@vsb.cz).}
\thanks{Tran Thien Thanh is with the Institute of Information Technology and Electrical-Electronics Engineering, Ho Chi Minh City University of Transport, Ho Chi Minh City, Vietnam (e-mail: thanh.tran@ut.edu.vn).}
}
\maketitle

\begin{abstract}\boldmath Simultaneously transmitting and reflecting reconfigurable intelligent surfaces (STAR-RISs) and rate-splitting multiple access (RSMA) offer full-space coverage and flexible interference management. However, covertness is hard to guarantee when the warden's channel is unknown to the transmitter.
We study covert rate maximization in a STAR-RIS-assisted RSMA massive multiple-input multiple-output downlink with all direct links blocked and an energy-detecting warden.
We prove that confining the covert precoder to the span of the cover precoders with a bounded norm guarantees covertness for every warden channel, including multi-antenna and colluding wardens, without warden channel state information.
On this basis, certified closed-form RSMA and non-orthogonal multiple access designs are derived, together with the public-rate requirement above which they fall short of the best certified designs.
Above this crossover, the design is cast as a one-step decision process and solved by a deep reinforcement learning (DRL) framework, in which a single actor trained with the differentiable system model as an analytic critic outputs the complete design through a certified decoder that makes every design covert.
Results on unseen channels show that the closed forms suffice at mild public-rate requirements and that the crossover shifts with the covertness budget as predicted.
Above the crossover, the actor outperforms all non-learned certified designs under the same single-thread computation time within a coherence block, with a gain that grows with the public-rate requirement up to $14.7\,\%$, and outperforms model-free DRL and classical reinforcement learning benchmarks.
\end{abstract}


\vspace{-0.25cm}

\section{Introduction}\label{sec:intro}

\subsection{Research Context}\label{sec:context}
\IEEEPARstart{T}{he} growing volume of sensitive data carried over the open wireless medium has made covert communication, which hides the very existence of a transmission from a warden, a stricter security paradigm than physical-layer secrecy \cite{bash2013limits}.
For a radiometer warden, covertness amounts to a lower bound on the minimum detection error probability (DEP), i.e., an upper bound on the ratio of the warden's received powers under its two hypotheses \cite{yan2019gaussian, vu2025performance}.
Since the warden is silent, the transmitter knows this ratio only statistically.

In parallel, simultaneously transmitting and reflecting reconfigurable intelligent surfaces (STAR-RISs) extend the coverage of conventional reconfigurable intelligent surfaces (RISs) to both sides of the surface \cite{mu2022simultaneously}, and rate-splitting multiple access (RSMA) manages multi-user interference through a common stream decoded by all users \cite{mao2018rate, tu2025active}.
Furthermore, a massive multiple-input multiple-output (MIMO) base station (BS) provides the array gain needed when covertness limits the transmit power.
We consider a deployment in which all direct links from the BS are blocked, so that the STAR-RIS provides the only path to the users.
The covert user and the warden then lie in the same transmission region and share the BS--STAR-RIS channel.
Without knowledge of the STAR-RIS--warden channel, the BS cannot steer the covert beam away from the warden, and covertness must therefore come from the STAR-RIS configuration and the stream structure.

The resulting design problem is nonconvex and must be solved within each coherence block.
Existing designs impose the DEP constraint either on its average over the warden's fading \cite{wu2021intelligent, xiao2024star, nguyen2025enhancing} or on the warden's instantaneous channel \cite{chang2024star, wang2026joint}.
The former leaves the realized channel unconstrained.
In Section~\ref{sec:results}, a design under this criterion violates the constraint on up to $40\,\%$ of the channels.
The latter requires the instantaneous channel of a passive warden, which is unavailable.
Both are solved per channel realization by alternating optimization (AO) or successive convex approximation (SCA).
Deep reinforcement learning (DRL) can replace this per-channel iteration with a trained neural network \cite{huang2020reconfigurable}. However, existing DRL-based covert designs penalize covertness violations in the reward and, like the optimization-based approaches, provide no covertness guarantee. This motivates the present work.

\subsection{Literature Overview}\label{sec:lit}
\emph{Covert STAR-RIS RSMA Designs:} Chang \etal~\cite{chang2024star} optimized the beamforming and the surface of a STAR-RIS-assisted multiple-input single-output RSMA covert downlink by AO, fractional programming (FP), and SCA, whereas Wang \etal~\cite{wang2026joint} jointly optimized the beamforming and the deployment of an active STAR-RIS for covert RSMA.
Both works constrain covertness on the warden's instantaneous channel.
Liang \etal~\cite{liang2025covert} and Yu \etal~\cite{yu2026covert} analyzed the DEP of an active STAR-RIS RSMA system with hardware impairments and of a terahertz STAR-RIS RSMA system with phase errors, respectively.
Nguyen \etal~\cite{nguyen2025enhancing} derived a closed-form power allocation for hybrid active--passive STAR-RIS RSMA, and Xiao \etal~\cite{xiao2024star} designed a STAR-RIS by AO and semidefinite relaxation (SDR) using only the warden's statistical channel state information (CSI), as Wu \etal~\cite{wu2021intelligent} did for a conventional RIS.
Beyond surfaces, covert millimeter-wave (mmWave) massive MIMO with hybrid beamforming was studied in \cite{bai2023covert, ci2021hybrid}, and hybrid beamforming for mmWave massive MIMO was developed in \cite{tu2025semi, tu2026hybrid, tu2026multi}.
Meanwhile, a public signal was used as a cover in non-orthogonal multiple access (NOMA) systems \cite{tao2020covert, vu2025performance, le2025active}, and our rate-splitting scheme follows the STAR-RIS RSMA framework of \cite{tu2025active, tu2026active}.

\emph{Learning-Based Designs:} DRL has been applied to surface-aided beamforming by the deep deterministic policy gradient (DDPG) \cite{huang2020reconfigurable}, to covert communication by twin delayed DDPG (TD3) \cite{yang2026intelligent} and proximal policy optimization (PPO) \cite{hieu2023joint}, and to STAR-RIS RSMA resource allocation by meta-learning DDPG \cite{amiri2025resource}.
Wang and Zhang \cite{wang2022intelligent} learned surface configurations from a codebook with a deep Q-network (DQN).
Closest to this work, Li \etal~\cite{li2026star} learned discretized powers, common-rate allocations, and a single surface amplitude for STAR-RIS-assisted covert RSMA by quantum reinforcement learning (RL). However, the covertness constraint in \cite{li2026star} again relies on the warden's instantaneous channel.

\emph{Research Gaps:} As summarized in Table~\ref{tab:novelty}, existing studies remain incomplete in three respects.
First, STAR-RIS RSMA covert designs impose covertness on the warden's average DEP or on its instantaneous channel, the latter even when only statistical warden CSI is assumed \cite{li2026star, chang2024star}.
To the best of our knowledge, no design that is covert for every warden channel without warden CSI has been reported.
Second, the conditions under which rate splitting is beneficial under guaranteed covertness have not been characterized.
Third, learning-based covert designs penalize covertness violations instead of excluding them, and none is evaluated against a non-learned optimizer on unseen channels, so that whether learning improves on non-learned certified designs remains open.

\begin{table*}[!t]
\centering
\caption{Comparison of This Work With Related Studies}
\label{tab:novelty}
\renewcommand{\arraystretch}{1.0}
\footnotesize
\resizebox{\ifdim\width>\textwidth\textwidth\else\width\fi}{!}{%
\begin{tabular}{|l|c|c|c|c|c|c|c|c|l|}
\hline
\textbf{Work} & \textbf{Year} & \textbf{STAR-RIS} & \textbf{\shortstack{RSMA/\\NOMA}} &
\textbf{\shortstack{Massive\\MIMO}} & \textbf{\shortstack{Hybrid\\beamforming}} & \textbf{\shortstack{No instantaneous\\warden CSI}} &
\textbf{\shortstack{Covert for every\\warden channel}} & \textbf{\shortstack{Learned\\policy}} & \textbf{Technique} \\
\hline
Tao \etal~\cite{tao2020covert} & 2020 & \xmark & NOMA & \xmark & \xmark & \cmark & \xmark & \xmark & DEP and outage analysis, power allocation \\
Huang \etal~\cite{huang2020reconfigurable} & 2020 & \xmark & \xmark & \xmark & \xmark & No warden & No warden & \cmark & DDPG \\
Wu \etal~\cite{wu2021intelligent} & 2021 & \xmark & \xmark & \xmark & \xmark & \cmark & \xmark & \xmark & Closed-form phases, 1-D search, CVX \\
Ci \etal~\cite{ci2021hybrid} & 2021 & \xmark & \xmark & \cmark & \cmark & \xmark & \xmark & \xmark & Semidefinite programming, SCA, alt.\ minimization \\
Wang and Zhang~\cite{wang2022intelligent} & 2022 & \xmark & \xmark & \xmark & \xmark & No warden & No warden & \cmark & DQN over a codebook \\
Bai \etal~\cite{bai2023covert} & 2023 & \xmark & \xmark & \cmark & \cmark & \cmark & \xmark & \xmark & Closed-form power, artificial noise \\
Hieu \etal~\cite{hieu2023joint} & 2023 & \xmark & RSMA   & \xmark & \xmark & \cmark & \xmark & \cmark & PPO \\
Xiao \etal~\cite{xiao2024star} & 2024 & \cmark & \xmark & \xmark & \xmark & \cmark & \xmark & \xmark & AO, SDR \\
Chang \etal~\cite{chang2024star} & 2024 & \cmark & RSMA   & \xmark & \xmark & \xmark & \xmark & \xmark & AO, FP, SCA \\
Vu \etal~\cite{vu2025performance} & 2025 & \xmark & NOMA   & \xmark & \xmark & \cmark & \xmark & \xmark & Closed form, CVX \\
Le \etal~\cite{le2025active} & 2025 & \xmark & NOMA & \xmark & \xmark & \cmark & \xmark & \xmark & Outage, secrecy, and DEP analysis \\
Amiri \etal~\cite{amiri2025resource} & 2025 & \cmark & RSMA   & \xmark & \xmark & No warden & No warden & \cmark & Meta-learning DDPG \\
Liang \etal~\cite{liang2025covert} & 2025 & \cmark & RSMA   & \xmark & \xmark & \cmark & \xmark & \xmark & Outage and DEP analysis \\
Nguyen \etal~\cite{nguyen2025enhancing} & 2025 & \cmark & RSMA   & \xmark & \xmark & \cmark & \xmark & \xmark & Closed-form power allocation, SCA \\
Yang \etal~\cite{yang2026intelligent} & 2026 & \xmark & \xmark & \xmark & \xmark & \xmark & \xmark & \cmark & TD3 \\
Yu \etal~\cite{yu2026covert} & 2026 & \cmark & RSMA   & \xmark & \xmark & \cmark & \xmark & \xmark & DEP analysis, power-allocation search \\
Wang \etal~\cite{wang2026joint} & 2026 & \cmark & RSMA   & \xmark & \xmark & \xmark & \xmark & \xmark & AO, FP, SCA, SDR \\
Li \etal~\cite{li2026star} & 2026 & \cmark & RSMA   & \xmark & \xmark & \xmark & \xmark & \cmark & Quantum RL \\
\hline
\textbf{This work} & -- & \cmark & RSMA, NOMA & \cmark & \cmark & \cmark & \cmark & \cmark & Certificate, closed forms, analytic-critic DRL \\
\hline
\end{tabular}}
\end{table*}

\subsection{Novelty and Contributions}
Motivated by these gaps, this work studies covert RSMA over a STAR-RIS-assisted massive MIMO downlink without warden CSI.
Here, the STAR-RIS provides the only path to both users, the RSMA common stream covers the covert stream, and the array gain compensates for the covertness-limited power.
The main contributions are summarized as follows.
\begin{enumerate}
\item \textbf{Covertness Certificate Without Warden CSI:} We prove that the DEP constraint of an energy-detecting warden holds for every warden channel, without warden CSI, if the covert precoder lies in the span of the cover precoders with a coefficient norm bounded by a threshold set by the covertness budget.
The guarantee extends to multi-antenna and colluding wardens and to imperfect legitimate CSI.
We further show that at most four phase-shifter radio-frequency (RF) chains realize any such design and derive the exact covert outage probability of designs outside this set.
\item \textbf{Certified Closed-Form Designs and Crossover Analysis:} We characterize when rate splitting is beneficial under the certificate and derive closed-form RSMA and NOMA designs that satisfy it.
In the high signal-to-noise ratio (SNR) limit, we predict the public-rate requirement above which they fall short of the best certified designs.
\item \textbf{Certified DRL Framework:} A deterministic actor outputs the complete design, namely the energy-splitting coefficients, phase shifts, and precoders, through a certified decoder that renders all its outputs covert.
It is trained by the deterministic policy gradient with the differentiable system model as an analytic critic.
One actor serves all operating points, and its output can initialize a short certified refinement.
\item \textbf{Extensive Simulations and Validation:} On unseen channels, all certified designs remain covert.
At demanding public-rate requirements, the learned design outperforms all non-learned certified designs under the same computation time within a coherence block, as well as model-free DRL and classical RL trained over certified designs.
\end{enumerate}
The technical novelty of this work lies in the covertness certificate and its integration with the closed-form and learned designs, and not in the individual learning tools.

\subsection{Organization and Notations}
The remainder of this paper is organized as follows.
Section~\ref{sec:model} presents the system model.
Section~\ref{sec:shield} develops the covertness certificate and the closed forms.
Section~\ref{sec:policy} introduces the DRL framework.
Section~\ref{sec:results} provides the numerical results.
Section~\ref{sec:conclusion} concludes the paper.

\emph{Notations:} Bold lowercase and uppercase letters denote vectors and matrices, respectively.
The operators $(\cdot)^{H}$, $(\cdot)^{T}$, $(\cdot)^{*}$, $(\cdot)^{\dagger}$, and $\operatorname{span}(\cdot)$ denote the conjugate transpose, the transpose, the conjugate, the pseudo-inverse, and the column space, respectively, and $[x]^{+}=\max\{x,0\}$.
Moreover, $\diag(\cdot)$ is a diagonal matrix, $\angle(\cdot)$ is the phase, and $\Exp\{\cdot\}$ is the expectation.
The functions $\mathrm{P}_{\Gamma}$ and $\mathrm{Q}_{\Gamma}$ are the regularized lower and upper incomplete Gamma functions, and $\sig(u)=(1+e^{-u})^{-1}$.
Furthermore, $\|\cdot\|$ is the Euclidean norm, $[\cdot]_{i}$ and $[\cdot]_{p,q}$ are the entries of a vector and a matrix, $\bs{I}_{N}$ is the $N\times N$ identity matrix, $\perp$ denotes orthogonality, and $\mathcal{CN}(\bs{\mu},\bs{\Sigma})$ is the complex Gaussian distribution with mean $\bs{\mu}$ and covariance $\bs{\Sigma}$.
Unless stated otherwise, the subscripts ${\rm n}$, ${\rm f}$, and ${\rm w}$ refer to the public user, the covert user, and the warden, respectively, and all rates are in bps/Hz.

\section{System Model and Problem Formulation}\label{sec:model}

\begin{figure}[!t]
\centering
\includegraphics[width=0.95\columnwidth]{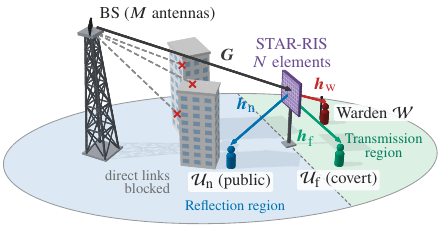}
\caption{System model of the covert STAR-RIS-assisted RSMA massive MIMO downlink, where all direct links are blocked.}
\label{fig:system}
\end{figure}

\subsection{Network and Channel Model}
We consider the downlink in Fig.~\ref{fig:system}, where a BS equipped with a uniform linear array of $M$ antennas serves a single-antenna public user $\mathcal{U}_{\rm n}$ and a single-antenna covert user $\mathcal{U}_{\rm f}$ through a passive STAR-RIS with $N=N_{\rm v}N_{\rm h}$ elements in $N_{\rm v}$ rows and $N_{\rm h}$ columns, in the presence of a passive single-antenna warden $\mathcal{W}$.
$\mathcal{U}_{\rm n}$ is located in the reflection region of the STAR-RIS, while $\mathcal{U}_{\rm f}$ and $\mathcal{W}$ are located in its transmission region.
We assume that all direct BS--node links are unavailable due to deep shadowing or blockage.
The STAR-RIS operates under the energy splitting protocol, in which each element divides the incident energy into reflected and transmitted components with independent phase shifts \cite{mu2022simultaneously}, and its reflection and transmission coefficient matrices are given by
\begin{equation}
\bs{\Theta}_{\rm R}=\diag\bigl(\sqrt{\beta_i^{\rm R}}e^{j\varphi_i}\bigr)_{i=1}^{N},\quad
\bs{\Theta}_{\rm T}=\diag\bigl(\sqrt{\beta_i^{\rm T}}e^{j\vartheta_i}\bigr)_{i=1}^{N},
\label{eq:starris}
\end{equation}
where $\beta_i^{\rm R}$ and $\beta_i^{\rm T}$ are the reflected and transmitted energy fractions of the $i$-th element with $\beta_i^{\rm R}+\beta_i^{\rm T}=1$ and $\beta_i^{\rm R}\in[0,1]$, $\varphi_i$ and $\vartheta_i$ are the corresponding phase shifts, $\bs{\Theta}_{\rm n}=\bs{\Theta}_{\rm R}$, and $\bs{\Theta}_{\rm f}=\bs{\Theta}_{\rm w}=\bs{\Theta}_{\rm T}$.
The BS--STAR-RIS channel $\bs{G}\in\Comp^{N\times M}$ is correlated at both ends, where the BS correlation follows the one-ring model as \cite{shiu2000fading}
\begin{equation}
[\bs{R}_{\rm BS}]_{p,q}=\frac{1}{2\Delta}\int_{-\Delta}^{\Delta}e^{\,jc_{0}(p-q)\sin(\bar\psi+x)}\,{\rm d}x,
\label{eq:onering}
\end{equation}
with $\bar\psi$ the mean angle of departure, $\Delta$ the angular spread, $c_{0}=2\pi d_{\rm e}/\lambda_{\rm c}$, $d_{\rm e}$ the BS antenna spacing, and $\lambda_{\rm c}$ the carrier wavelength.
The STAR-RIS correlation follows the separable counterpart of \eqref{eq:onering}.
The STAR-RIS--node channels $\bs{h}_{k}\in\Comp^{N}$ are independent and identically distributed (i.i.d.) Rayleigh (rich scattering without a line of sight) with $\Exp\{|h_{k,i}|^{2}\}=d_{k}^{-\eta}$, where $d_{k}$ is the STAR-RIS--node $k$ distance, $\eta$ is the path-loss exponent \cite{tu2025active}, and the reference gain of Table~\ref{tab:params} is omitted.
Accordingly, the cascaded BS--node $k$ channel is
\begin{equation}
\bs{a}_{k}^{H}=\bs{h}_{k}^{H}\bs{\Theta}_{k}\bs{G}\in\Comp^{1\times M},\qquad k\in\{{\rm n},{\rm f},{\rm w}\}.
\label{eq:rows}
\end{equation}
Note that $\bs{a}_{\rm f}$ and $\bs{a}_{\rm w}$ share the factor $\bs{\Theta}_{\rm T}\bs{G}$ and differ only in $\bs{h}_{\rm f}$ and $\bs{h}_{\rm w}$.

\subsection{RSMA Transmission and Achievable Rates}\label{sec:rsma}
The covert user $\mathcal{U}_{\rm f}$ receives regular data and a covert message.
In one-layer RSMA \cite{mao2018rate, tu2025active}, each message is split into a common and a private part, where the common parts are jointly encoded into $s_{\rm c}$, and the private part of $\mathcal{U}_{\rm n}$ and the covert message of $\mathcal{U}_{\rm f}$ are encoded into $s_{\rm n}$ and $s_{\rm f}$, respectively.
The BS transmits
\begin{equation}
\bs{x}=\sqrt{P}\left(\bs{w}_{\rm c}s_{\rm c}+\bs{w}_{\rm n}s_{\rm n}+\bs{w}_{\rm f}s_{\rm f}\right),
\label{eq:signal}
\end{equation}
where the symbols are unit-power Gaussian, $\bs{w}_{m}$ is the precoder of $s_{m}$ with $\sum_{m\in\{{\rm c},{\rm n},{\rm f}\}}\|\bs{w}_{m}\|^{2}=1$, and $P\le P_{S}$ is the transmit power for a power budget $P_{S}$.
Node $k$ receives $y_{k}=\bs{a}_{k}^{H}\bs{x}+n_{k}$ with noise $n_{k}\sim\mathcal{CN}(0,\sigma_{k}^{2})$ and $\sigma_{k}^{2}=\sigma^{2}$ for all $k$.
Each user first decodes and removes $s_{\rm c}$ by successive interference cancellation (SIC) before decoding its private stream.
With $p_{k,m}=|\bs{a}_{k}^{H}\bs{w}_{m}|^{2}$, the signal-to-interference-plus-noise ratios (SINRs) of $s_{\rm c}$ at user $k$, of $s_{\rm n}$ at $\mathcal{U}_{\rm n}$, and of $s_{\rm f}$ at $\mathcal{U}_{\rm f}$ are defined, respectively, as
\begin{align}
\gamma_{{\rm c},k}&=\frac{Pp_{k,{\rm c}}}{P(p_{k,{\rm n}}+p_{k,{\rm f}})+\sigma^{2}},\quad k\in\{{\rm n},{\rm f}\},\label{eq:sinr}\\
\gamma_{\rm n}&=\frac{Pp_{{\rm n},{\rm n}}}{Pp_{{\rm n},{\rm f}}+\sigma^{2}},\label{eq:sinr_n}\\
\gamma_{\rm f}&=\frac{Pp_{{\rm f},{\rm f}}}{Pp_{{\rm f},{\rm n}}+\sigma^{2}},\label{eq:sinr_f}
\end{align}
and the common rate is given by $R_{\rm c}=\min_{k}\log_{2}(1+\gamma_{{\rm c},k})$.
The covert user is allocated a fraction $\tau\in(0,1)$ of $R_{\rm c}$ and the public user the remaining $(1-\tau)R_{\rm c}$, where $\tau$ is an optimization variable, so that the public user's rate and the covert rate are
\begin{equation}
R_{\rm n}=(1-\tau)R_{\rm c}+\log_{2}(1+\gamma_{\rm n}),\,\, R_{\rm f}^{\rm cov}=\log_{2}(1+\gamma_{\rm f}).
\label{eq:rates}
\end{equation}
Only $R_{\rm f}^{\rm cov}$ counts as covert, since the common share $\tau R_{\rm c}$ is sent under both hypotheses (Section~\ref{sec:warden}) and decoded by the public user.
Following \cite{tu2025active}, the rate requirements are given by
\begin{equation}
\min\{\tau,1-\tau\}R_{\rm c}\ge R_{\rm c}^{\rm th},\,\, R_{\rm n}\ge R_{\rm th},\,\, R_{\rm f}^{\rm cov}\ge R_{\rm f}^{\rm th},
\label{eq:qos}
\end{equation}
where $R_{\rm c}^{\rm th}$ is the minimum common rate per user, and $R_{\rm th}$ and $R_{\rm f}^{\rm th}$ are the minimum public-user total rate, referred to as the public-rate requirement or public requirement for short, and covert rate, respectively.

\subsection{Warden Detection and Covertness Constraint}\label{sec:warden}
The warden $\mathcal{W}$ aims to detect whether $s_{\rm f}$ is transmitted.
Since $s_{\rm c}$ and $s_{\rm n}$ are transmitted with the same precoders and power in both cases, the warden faces the following binary hypothesis test \cite{tao2020covert, vu2025performance}:
\begin{equation}
\begin{aligned}
\mathcal{H}_{0}&:\ \bs{x}=\sqrt{P}(\bs{w}_{\rm c}s_{\rm c}+\bs{w}_{\rm n}s_{\rm n}),\\
\mathcal{H}_{1}&:\ \bs{x}=\sqrt{P}(\bs{w}_{\rm c}s_{\rm c}+\bs{w}_{\rm n}s_{\rm n}+\bs{w}_{\rm f}s_{\rm f}),
\end{aligned}
\label{eq:hypo}
\end{equation}
and compares the energy of $n$ received samples with a threshold.
For a given warden channel, the average sample power is $\mu_{0}=P(p_{{\rm w},{\rm c}}+p_{{\rm w},{\rm n}})+\sigma_{\rm w}^{2}$ under $\mathcal{H}_{0}$ and $\mu_{1}=\mu_{0}+Pp_{{\rm w},{\rm f}}$ under $\mathcal{H}_{1}$.
With the optimal threshold, the minimum DEP depends only on $\rho=\mu_{1}/\mu_{0}$ and is given by \cite{yan2019gaussian, vu2025performance}
\begin{equation}
\mathrm{DEP}^{\star}(n,\rho)=\mathrm{Q}_{\Gamma}\Bigl(n,\frac{n\rho\ln\rho}{\rho-1}\Bigr)+\mathrm{P}_{\Gamma}\Bigl(n,\frac{n\ln\rho}{\rho-1}\Bigr),
\label{eq:depmin}
\end{equation}
which decreases in $\rho$.
Hence, for a covertness budget $\epsilon\in(0,1)$, $\mathrm{DEP}^{\star}\ge1-\epsilon$ is equivalent to
\begin{equation}
\rho=1+\frac{Pp_{{\rm w},{\rm f}}}{P(p_{{\rm w},{\rm c}}+p_{{\rm w},{\rm n}})+\sigma_{\rm w}^{2}}\ \le\ \rho^{\star},
\label{eq:covertness}
\end{equation}
where $\rho^{\star}$ is the solution of $\mathrm{DEP}^{\star}(n,\rho)=1-\epsilon$, and $\varrho\triangleq\rho^{\star}-1$.

\begin{assumption}[Warden and CSI]\label{ass:warden}
The warden is a static single-antenna radiometer with perfect knowledge of its realized channel and noise power.
It uses the optimal threshold and does not decode the streams.\footnote{This model follows \cite{vu2025performance, le2025active}. Proposition~\ref{prop:multi} covers multi-antenna and colluding wardens.}
The BS has perfect CSI of $\bs{G}$ and of the users' cascaded channels.
As the warden sends no pilots, the BS knows only the statistics $\bs{R}_{h}=\Exp\{\bs{h}_{\rm w}\bs{h}_{\rm w}^{H}\}=d_{\rm w}^{-\eta}\bs{I}_{N}$ of $\bs{h}_{\rm w}$ (Section~\ref{sec:res-setup}).
\end{assumption}

The radiometer warden of Assumption~\ref{ass:warden} \cite{yan2019gaussian, tao2020covert, vu2025performance} measures only the total received energy, since it lacks the codebooks and frame synchronization needed to decode the common stream.
Its knowledge of the realized channel and noise power is the worst case under energy detection.
A warden that decodes and cancels the public streams \cite{ahmadi2026square, wang2026joint} is a different threat model outside the certificate, which is examined numerically in Section~\ref{sec:res-cov}.

Under Assumption~\ref{ass:warden}, $\rho$ in \eqref{eq:covertness} depends on the realized warden channel $\bs{a}_{\rm w}$, which is unknown at the BS, and is written as $\rho(\bs{a}_{\rm w})$.
A conventional design therefore imposes \eqref{eq:covertness} on the believed (average) warden powers $\hat p_{{\rm w},m}=\bs{w}_{m}^{H}\bs{C}\bs{w}_{m}$, with $\bs{C}=\bs{C}(\bs{\Theta}_{\rm w})=\bs{G}^{H}\bs{\Theta}_{\rm w}^{H}\bs{R}_{h}\bs{\Theta}_{\rm w}\bs{G}$, and transmits with $P=\min\{P^{\star},P_{S}\}$, where
\begin{equation}
P^{\star}=\frac{\sigma_{\rm w}^{2}\varrho}{\hat p_{{\rm w},{\rm f}}-\varrho(\hat p_{{\rm w},{\rm c}}+\hat p_{{\rm w},{\rm n}})},
\label{eq:power}
\end{equation}
is the largest power meeting this constraint and $P^{\star}=\infty$ if the denominator of \eqref{eq:power} is nonpositive.
However, such a design may still violate \eqref{eq:covertness} on the realized warden channel.

\subsection{Problem Formulation}\label{sec:problem}
In this work, we aim to maximize the covert rate by jointly optimizing the transmit power, the precoders, the STAR-RIS coefficients, and the common-rate split $\tau$.
Let $\bs{\xi}=\{P,\bs{w}_{\rm c},\bs{w}_{\rm n},\bs{w}_{\rm f},\bs{\beta}^{\rm R},\bs{\varphi},\bs{\vartheta},\tau\}$ denote the set of optimization variables and $\bs{H}=\{\bs{G},\bs{h}_{\rm n},\bs{h}_{\rm f}\}$ the set of legitimate channels, where $\bs{\beta}^{\rm R}$, $\bs{\varphi}$, and $\bs{\vartheta}$ collect the per-element coefficients.
For a given channel realization, the covert rate maximization problem is formulated as
\begin{subequations}\label{eq:P1}
\begin{align}
(\mathrm{P1}):\ \max_{\bs{\xi}}\ \ & R_{\rm f}^{\rm cov}(\bs{\xi};\bs{H}) \label{eq:P1obj}\\
\text{s.t.}\ \ & \rho(\bs{a}_{\rm w})\le\rho^{\star}, \label{eq:P1cov}\\
& \eqref{eq:qos},\quad P\le P_{S}, \label{eq:P1qos}\\
& \textstyle\sum_{m}\|\bs{w}_{m}\|^{2}=1,\ \ \tau\in(0,1), \label{eq:P1pow}\\
& \beta_i^{\rm R}+\beta_i^{\rm T}=1,\ \ \beta_i^{\rm R}\in[0,1],\ \forall i. \label{eq:P1star}
\end{align}
\end{subequations}
Constraint \eqref{eq:P1cov} imposes covertness on the realized warden channel, \eqref{eq:P1qos} collects the rate and power limits, \eqref{eq:P1pow} is the precoder normalization, and \eqref{eq:P1star} is the per-element energy conservation.
(P1) is nonconvex in the precoders and the STAR-RIS coefficients, and constraint \eqref{eq:P1cov} cannot be evaluated at the BS.
Section~\ref{sec:shield} replaces \eqref{eq:P1cov} by a constraint that does not involve the warden channel.

\section{Covertness Analysis and Certified Closed Forms}\label{sec:shield}

\subsection{Certified Covertness by Common-Stream Shielding}
Let $\bs{W}=[\bs{w}_{\rm c},\bs{w}_{\rm n}]$ collect the cover precoders.
For any warden channel $\bs{a}\in\Comp^{M}$, $\rho$ in \eqref{eq:covertness} becomes
\begin{equation}
\rho(\bs{a})=1+\frac{P|\bs{a}^{H}\bs{w}_{\rm f}|^{2}}{P\|\bs{W}^{H}\bs{a}\|^{2}+\sigma_{\rm w}^{2}}.
\label{eq:rho_a}
\end{equation}

\begin{proposition}[Certified covertness]\label{prop:shield}
Consider the warden of Assumption~\ref{ass:warden} and $P>0$.
(\textit{i}) If
\begin{equation}
\bs{w}_{\rm f}=\bs{W}\bs{b}\ \ \text{for some}\ \bs{b}\in\Comp^{2}\ \text{with}\ \|\bs{b}\|^{2}\le\varrho,
\label{eq:shield}
\end{equation}
then $\rho(\bs{a})\le\rho^{\star}$ for every $\bs{a}\in\Comp^{M}$ and every warden noise power $\sigma_{\rm w}^{2}>0$.
(\textit{ii}) Conversely, if $\sigma_{\rm w}^{2}$ does not depend on $\bs{a}$ and $\bs{a}$ may be any vector of a subspace $\mathcal{S}$ that contains $\bs{w}_{\rm f}$ and the columns of $\bs{W}$, then $\sup_{\bs{a}\in\mathcal{S}}\rho(\bs{a})\le\rho^{\star}$ implies \eqref{eq:shield} with $\bs{b}=\bs{W}^{\dagger}\bs{w}_{\rm f}$.
For the surface-side warden, $\mathcal{S}=\operatorname{span}(\bs{G}^{H})$.
\end{proposition}
\begin{IEEEproof}
See Appendix~\ref{app:shield}.
\end{IEEEproof}

Proposition~\ref{prop:shield} extends the single-antenna cover-based NOMA designs of \cite{tao2020covert, vu2025performance}, where the covert signal is colinear with the cover and $\rho\le1+P_{\rm f}/P_{\rm c}$ for any warden channel, with $P_{\rm f}$ and $P_{\rm c}$ the covert and cover powers, to multi-antenna precoders.

\begin{remark}\label{rem:shield}
A design satisfying \eqref{eq:shield} is called \emph{shielded}, or equivalently certified.
It needs no knowledge of the warden's channel, statistics, or noise power, and thus operates at full power $P_{S}$.
The covert user removes $s_{\rm c}$ by SIC before decoding $s_{\rm f}$, but the warden's energy statistic cannot separate the streams.
For given cover precoders, the covert power is maximized by $\bs{b}^{\star}=\sqrt{\varrho}\,\bs{W}^{H}\bs{a}_{\rm f}/\|\bs{W}^{H}\bs{a}_{\rm f}\|$, so that the covert rate grows with the cover power received at the covert user, which depends on both the surface and the precoders.
\end{remark}

Consider next a warden with $N_{\rm w}$ antennas, or colluding wardens that jointly process $N_{\rm w}$ antennas in total.
Let $\bs{A}\in\Comp^{M\times N_{\rm w}}$ and $\bs{\Sigma}\succ\bs{0}$ be its channel and positive-definite noise covariance, and assume it knows both hypothesis covariances $\bs{\Sigma}_{0}=P\bs{A}^{H}\bs{W}\bs{W}^{H}\bs{A}+\bs{\Sigma}$ and $\bs{\Sigma}_{1}=\bs{\Sigma}_{0}+P\bs{A}^{H}\bs{w}_{\rm f}\bs{w}_{\rm f}^{H}\bs{A}$.

\begin{proposition}[Multi-antenna and colluding wardens]\label{prop:multi}
If \eqref{eq:shield} holds and the streams are Gaussian, the minimum DEP over all detectors that do not decode the streams satisfies $\mathrm{DEP}_{\min}=\mathrm{DEP}^{\star}(n,1+\zeta)\ge1-\epsilon$, where $\zeta=P\bs{w}_{\rm f}^{H}\bs{A}\bs{\Sigma}_{0}^{-1}\bs{A}^{H}\bs{w}_{\rm f}\le\|\bs{b}\|^{2}\le\varrho$, for every $\bs{A}$, $N_{\rm w}$, and $\bs{\Sigma}$.
\end{proposition}
\begin{IEEEproof}
See Appendix~\ref{app:shield}.
\end{IEEEproof}

Since Propositions~\ref{prop:shield} and~\ref{prop:multi} hold for arbitrary $\bs{W}$ and surface coefficients, a shielded design remains covert under imperfect legitimate CSI, surface phase errors, and coefficient quantization, which affect only the rates.
The colinear form is $\bs{w}_{\rm f}=c\,\bs{w}_{\rm c}$ with $|c|^{2}\le\varrho$, and the nulled form also cancels the covert stream at the public user such that
\begin{equation}
\bs{w}_{\rm f}=b_{0}\bigl(\bs{w}_{\rm c}-\nu\bs{w}_{\rm n}\bigr),\quad
\nu=\frac{\bs{a}_{\rm n}^{H}\bs{w}_{\rm c}}{\bs{a}_{\rm n}^{H}\bs{w}_{\rm n}},\quad
|b_{0}|^{2}(1+|\nu|^{2})\le\varrho.
\label{eq:nullshield}
\end{equation}
Let (P1-S) and (P1-S$'$) denote (P1) at $P=P_{S}$ with \eqref{eq:P1cov} replaced by the colinear and the nulled form \eqref{eq:nullshield}, respectively.

\subsection{Covert Outage, Hybrid Realization, and RSMA Versus NOMA}
For a non-shielded design, let $\bs{h}_{\rm w}\sim\mathcal{CN}(\bs{0},\bs{R}_{h})$, $\bs{q}_{m}=\sqrt{P}\bs{\Theta}_{\rm w}\bs{G}\bs{w}_{m}$ and
\begin{equation}
\bs{M}=\bs{q}_{\rm f}\bs{q}_{\rm f}^{H}-\varrho\bigl(\bs{q}_{\rm c}\bs{q}_{\rm c}^{H}+\bs{q}_{\rm n}\bs{q}_{\rm n}^{H}\bigr),
\label{eq:Mmatrix}
\end{equation}
so that \eqref{eq:covertness} is violated if and only if $\bs{h}_{\rm w}^{H}\bs{M}\bs{h}_{\rm w}>\varrho\sigma_{\rm w}^{2}$.

\begin{proposition}[Exact covert outage]\label{prop:outage}
Let $\lambda_{1}\ge\cdots\ge\lambda_{N}$ be the eigenvalues of $\bs{R}_{h}^{1/2}\bs{M}\bs{R}_{h}^{1/2}$.
At most one is positive, and
\begin{equation}
\begin{aligned}
P_{\rm out}&\triangleq\Pr\{\mathrm{DEP}^{\star}<1-\epsilon\}\\
&=\begin{cases}
e^{-\varrho\sigma_{\rm w}^{2}/\lambda_{1}}\prod_{j=2}^{N}\frac{\lambda_{1}}{\lambda_{1}-\lambda_{j}}, & \lambda_{1}>0,\\
0, & \lambda_{1}\le0.
\end{cases}
\end{aligned}
\label{eq:pout}
\end{equation}
\end{proposition}
\begin{IEEEproof}
See Appendix~\ref{app:outage}.
\end{IEEEproof}

A shielded design has $\lambda_{1}\le0$ and hence $P_{\rm out}=0$, whereas a constraint on the believed powers in \eqref{eq:power} does not control $P_{\rm out}$ on the realized channel, as formalized next.

\begin{corollary}[Outage floor under statistical CSI]\label{cor:floor}
The power constraint $P\le P^{\star}$ in \eqref{eq:power}, which relies on the believed powers, is equivalent to $\sum_{j}\lambda_{j}\le\varrho\sigma_{\rm w}^{2}$.
Hence, any design operated at $P=P^{\star}<P_{S}$ has $P_{\rm out}\ge e^{-1}\approx0.37$, regardless of its precoders and surface coefficients.
\end{corollary}
\begin{IEEEproof}
Since $P\hat p_{{\rm w},m}=\bs{q}_{m}^{H}\bs{R}_{h}\bs{q}_{m}$, we have $P(\hat p_{{\rm w},{\rm f}}-\varrho(\hat p_{{\rm w},{\rm c}}+\hat p_{{\rm w},{\rm n}}))=\operatorname{tr}(\bs{R}_{h}^{1/2}\bs{M}\bs{R}_{h}^{1/2})=\sum_{j}\lambda_{j}$, which \eqref{eq:power} sets to $\varrho\sigma_{\rm w}^{2}$ when $P=P^{\star}$.
Then $\lambda_{1}=\varrho\sigma_{\rm w}^{2}+\sum_{j\ge2}|\lambda_{j}|$, and applying $1+x\le e^{x}$ to \eqref{eq:pout} gives $P_{\rm out}\ge\exp\bigl(-(\varrho\sigma_{\rm w}^{2}+\sum_{j\ge2}|\lambda_{j}|)/\lambda_{1}\bigr)=e^{-1}$.
\end{IEEEproof}

\begin{corollary}[Hybrid realization]\label{cor:hybrid}
Let $\bs{W}_{\rm all}=[\bs{w}_{\rm c},\bs{w}_{\rm n},\bs{w}_{\rm f}]$ satisfy \eqref{eq:shield} and be realized as $\bs{F}\bs{B}$ with a constant-modulus analog stage $\bs{F}\in\Comp^{M\times N_{\rm RF}}$ of $N_{\rm RF}$ RF chains.
(\textit{i}) For every $\bs{F}$, the least-squares digital stage $\bs{B}=\bs{F}^{\dagger}\bs{W}_{\rm all}$ again satisfies \eqref{eq:shield}.
(\textit{ii}) $\operatorname{rank}\bs{W}_{\rm all}=\operatorname{rank}\bs{W}\le2$.
Hence, at most four phase-shifter RF chains are required to realize $\bs{W}_{\rm all}$ exactly, regardless of $M$.
\end{corollary}
\begin{IEEEproof}
See Appendix~\ref{app:outage}.
\end{IEEEproof}

Under \eqref{eq:shield}, the public user must decode both cover streams.
Let $r=R_{\rm c}^{\rm th}$ and let (P1-N) denote the shielded NOMA design, in which the common stream carries only the public message ($\tau=0$, $\bs{w}_{\rm n}=\bs{0}$, $R_{\rm c}\ge R_{\rm th}$, and no minimum common-rate share for the covert user in \eqref{eq:qos}) and the covert stream is colinear.

\begin{proposition}[RSMA versus NOMA]\label{prop:rsma}
For shielded designs at $P=P_{S}$ and any channel realization, the following statements hold.
(\textit{i}) The optimum of (P1-N) is at least that of (P1-S) with $\bs{w}_{\rm n}=\bs{0}$, and at least that of (P1-S) if $R_{\rm th}\le2r$.
(\textit{ii}) Feasibility of (P1-N) requires $|c|^{2}<1/(2^{R_{\rm th}}-1)$, and, for $R_{\rm th}\ge r$, feasibility of (P1-S) requires $|c|^{2}<1/\bigl(2^{R_{\rm th}-r}(2^{2r}-1)\bigr)$.
(\textit{iii}) In (P1-S$'$), $p_{{\rm n},{\rm f}}=0$ by construction, and feasibility requires $|b_{0}|^{2}<\varrho 2^{-2r}$, a bound independent of $R_{\rm th}$.
\end{proposition}

\begin{IEEEproof}
See Appendix~\ref{app:rsma}.
\end{IEEEproof}

\begin{remark}[Predicted crossover]\label{rem:crossover}
In the high-SNR limit, if each bound in Proposition~\ref{prop:rsma} is approached with the same cover power at the covert user, (P1-S$'$) outperforms NOMA when
\begin{equation}
R_{\rm th}>\log_{2}\bigl(1+2^{2r}/\varrho \bigr).
\label{eq:crossover}
\end{equation}
Condition \eqref{eq:crossover} serves as a design guideline rather than a bound, and its accuracy is examined numerically in Section~\ref{sec:results}.
\end{remark}

\begin{remark}\label{rem:rsma}
By Proposition~\ref{prop:rsma}(\textit{i}), rate splitting under shielding is beneficial only through the public user's private stream, which acts as a second cover and provides the degree of freedom to cancel the covert stream at the public user, and only when $R_{\rm th}>2r$.
\end{remark}

\subsection{Certified Closed-Form Designs}\label{sec:cf}
\begin{corollary}[Colinear design without a public private stream]\label{cor:cf}
Let $\bs{w}_{\rm n}=\bs{0}$, $\bs{w}_{\rm f}=c\,\bs{w}_{\rm c}$ and $P=P_{S}$, and fix the surface.
(\textit{i}) The covert rate satisfies
\begin{equation}
R_{\rm f}^{\rm cov}\le\log_{2}\Bigl(1+\frac{|c|^{2}}{1+|c|^{2}}\,\frac{P_{S}\|\bs{a}_{\rm f}\|^{2}}{\sigma^{2}}\Bigr),
\label{eq:cfbound}
\end{equation}
with equality for $c\neq0$ if and only if $\bs{w}_{\rm c}\propto\bs{a}_{\rm f}$, and the bound is increasing in $|c|^{2}$.
(\textit{ii}) Every feasible design has $|c|^{2}<1/(2^{R_{\rm th}+r}-1)$.
\end{corollary}
\begin{IEEEproof}
See Appendix~\ref{app:rsma}.
\end{IEEEproof}

By Corollary~\ref{cor:cf}, the colinear design reduces to optimizing the surface, and the colinear closed form is obtained in four steps.
First, the transmission phases follow the covert user's cascaded channel, $\vartheta_i=-\angle(h_{{\rm f},i}^{*}[\bs{G}\bs{w}]_{i})$, with $\bs{\vartheta}$ and $\bs{w}$ updated alternately, where $\bs{w}$ is the matched filter of $\bs{a}_{\rm f}$.
Second, we set $\bs{w}_{\rm c}\propto\bs{a}_{\rm f}$ and $\bs{w}_{\rm f}=c\,\bs{w}_{\rm c}$.
Third, the reflection phases follow the public user's cascaded channel given $\bs{w}_{\rm c}$.
Finally, $\beta^{\rm R}$ is the smallest common reflected fraction that admits a feasible $\tau$, and $|c|^{2}$ is the largest feasible value below the bound in Corollary~\ref{cor:cf}(\textit{ii}).
The NOMA closed form is the same design with $\tau=0$ and the looser cap of Proposition~\ref{prop:rsma}(\textit{ii}).
Since the bound on $|c|^{2}$ decreases with $R_{\rm th}$, the remaining closed forms carry part of the public message by a private stream zero-forced at the covert user,
\begin{equation}
\bs{w}_{\rm n}\propto\bs{P}^{\perp}_{\rm f}\bs{a}_{\rm n},\quad \bs{P}^{\perp}_{\rm f}=\bs{I}_{M}-\bs{u}_{\rm f}\bs{u}_{\rm f}^{H},\quad
\bs{u}_{\rm f}=\bs{a}_{\rm f}/\|\bs{a}_{\rm f}\|,
\label{eq:zfpub}
\end{equation}
where $\bs{P}^{\perp}_{\rm f}$ is the orthogonal projector onto the orthogonal complement of $\bs{a}_{\rm f}$.
Above the crossover \eqref{eq:crossover}, the nulled closed form uses \eqref{eq:nullshield} with the largest admissible $|b_{0}|^{2}$ and the private stream \eqref{eq:zfpub}.
Setting the common rate to its minimum and the common-to-private power ratio at the public user analytically close to $2^{2r}-1$ approaches the bound in Proposition~\ref{prop:rsma}(\textit{iii}).
Since $\bs{G}$ is common to both users and correlated, the reflection phases follow the part of the public user's cascaded channel left after zero-forcing,
\begin{equation}
\varphi_i=-\angle\bigl(h_{{\rm n},i}^{*}[\bs{G}\bs{P}^{\perp}_{\rm f}\bs{v}]_{i}\bigr),
\label{eq:zfaware}
\end{equation}
alternated with $\bs{v}$ as the matched filter of the resulting row.
The element-partition closed form keeps the colinear shielding and the private stream \eqref{eq:zfpub} but uses two energy-splitting levels, where the $K$ elements with the largest $|h_{{\rm n},i}|^{2}/|h_{{\rm f},i}|^{2}$ reflect a fraction $\beta_{\rm hi}$ and the others a fraction $\beta_{\rm lo}$.
None of these closed forms requires training or warden CSI, and their covertness follows from Proposition~\ref{prop:shield}.
Only the scalars $\beta^{\rm R}$, $\tau$, the shielding coefficients, the rate margins, $K$, $\beta_{\rm hi}$, and $\beta_{\rm lo}$ are found by grid or bisection search (Section~\ref{sec:res-setup}).
Per channel realization, we refer to the closed form with the highest covert rate as the \emph{best closed form}.

\section{Proposed DRL Framework}\label{sec:policy}

\subsection{Decision Process and Motivation for Learning}\label{sec:mdp}
Since (P1) must be solved in each coherence block for a fixed geometry, we seek a policy $\pi$ that maps channel observations to a design by solving
\begin{equation}
(\mathrm{P2})\quad \max_{\pi}\ \Exp_{\bs{H}}\bigl\{S\bigl(\pi(\bs{o}(\bs{H}));\bs{H}\bigr)\bigr\},
\label{eq:P2}
\end{equation}
where $\bs{o}(\bs{H})$ uses only quantities known at the transmitter and $S$ is the covert rate if all (P1) constraints hold and zero otherwise.
Since the channels are independent across blocks, (P2) is a contextual bandit, i.e., the single-step case of RL \cite{sutton2018reinforcement}, with state $\bs{s}_{t}=\bs{o}(\bs{H}_{t})$, action $\bs{\xi}_{t}$, and reward $S$.

\emph{Motivation for Learning:} Under stringent rate requirements and limited power, the best certified designs found by the per-channel solver of Section~\ref{sec:res-setup} share a common structure.
The common stream stays at the users' minimum shares, the public user is served through its own private stream, and the surface energy split is non-uniform.
The closed forms of Section~\ref{sec:shield} capture this structure only partially, with at most two energy-split levels, while the solver needs thousands of gradient steps per channel, which is prohibitive within a coherence block.
We therefore train an actor, in the design space restricted by Section~\ref{sec:shield}, to output the design of each channel in a single forward pass.

\subsection{Certified Actor Architecture}\label{sec:arch}

\begin{figure*}[!t]
\centering
\includegraphics[width=0.8\textwidth]{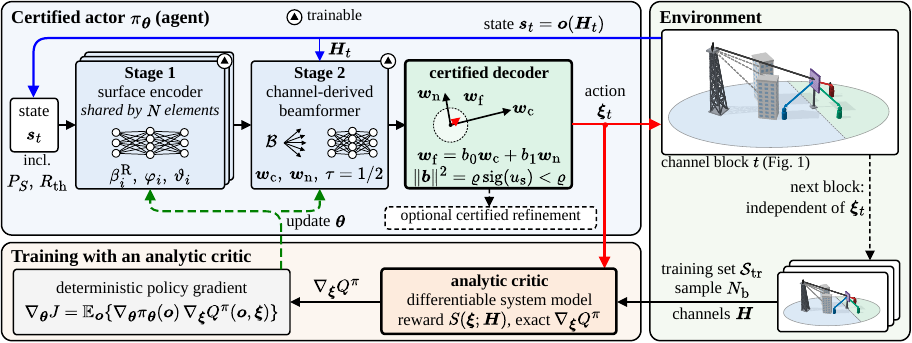}
\caption{Proposed certified DRL framework with an analytic critic.}
\label{fig:architecture}
\end{figure*}

In the framework of Fig.~\ref{fig:architecture}, the state comprises $37$ features per element and $13$ context features, including the transmit power and the public requirement.
The element features are the log-magnitude, cosine, and sine of $h_{k,i}^{*}[\bs{G}\bs{v}]_{i}$ for both users, with $\bs{v}$ equal to each user's matched filter $\bs{G}^{H}\bs{h}_{k}$ (entered twice, in place of the absent direct links) and to the dominant right singular vector of $\bs{G}$.
They also include the same quantities for the leading eigenvector of $\bs{C}(\bs{I}_{N})$, which represents the warden's believed direction, as well as $\ln|h_{{\rm n},i}|$, $\ln|h_{{\rm f},i}|$, $\ln[\bs{R}_{h}]_{i,i}$, and the mean log-magnitude of the $i$-th row of $\bs{G}$.
The warden's realized channel is not used.

\emph{Stage 1 (Surface Encoder):} An encoder shared by the $N$ elements outputs the energy split $\beta_i^{\rm R}=\sig(\cdot)$ of the $i$-th element and a correction $\Delta\varphi_i$ to its coherent alignment phase:
\begin{equation}
\varphi_i=-\angle\bigl(h_{{\rm n},i}^{*}[\bs{G}\bs{v}_{\rm n}]_{i}\bigr)+\Delta\varphi_i,
\label{eq:align}
\end{equation}
where $\bs{v}_{\rm n}=\bs{G}^{H}\bs{h}_{\rm n}/\|\bs{G}^{H}\bs{h}_{\rm n}\|$, and likewise $\vartheta_i$ for the covert user.

\emph{Stage 2 (Channel-Derived Beamformer):} Given the surface, a second head outputs the power split and complex coefficients over the basis
\begin{equation}
\mathcal{B}=\bigl\{\bs{a}_{\rm f},\ \bs{a}_{\rm n},\ \bs{C}^{-1}\bs{a}_{\rm f},\ \bs{C}^{-1}\bs{a}_{\rm n},\ \bs{a}_{\rm f}^{\perp\bs{a}_{\rm n}}\bigr\},
\label{eq:basis}
\end{equation}
where $\bs{C}^{-1}$ denotes the regularized inverse $(\bs{C}+10^{-6}\bar{c}\,\bs{I}_{M})^{-1}$ with $\bar{c}=\operatorname{tr}(\bs{C})/M$, which whitens the matched filters by the warden statistics.
The regularization is needed because $\bs{C}$ has rank at most $\operatorname{rank}(\bs{G})$.
The last element of $\mathcal{B}$ is $\bs{a}_{\rm f}^{\perp\bs{a}_{\rm n}}=(\bs{I}_{M}-\bs{a}_{\rm n}\bs{a}_{\rm n}^{H}/\|\bs{a}_{\rm n}\|^{2})\bs{a}_{\rm f}$, i.e., the component of the covert filter orthogonal to the public one.
The common share is fixed to $\tau=1/2$, which meets both common-rate floors in \eqref{eq:qos} with the smallest common rate.

\emph{Certified Decoder:} The covert precoder lies in the span of the cover precoders,
\begin{equation}
\bs{w}_{\rm f}=b_{0}\bs{w}_{\rm c}+b_{1}\bs{w}_{\rm n},\qquad \|\bs{b}\|^{2}=\varrho\,\sig(u_{\rm s}),
\label{eq:decoder}
\end{equation}
where $\bs{b}=[b_{0},b_{1}]^{T}$ and $u_{\rm s}$ is a scalar network output.
This ensures $\|\bs{b}\|^{2}<\varrho$ for any network output and hence, by Proposition~\ref{prop:shield}, covertness for every warden channel.
Both closed-form shields lie in the closure of the decoder's output set.
The encoders and heads are multilayer perceptrons with sigmoid-weighted linear unit activations.
Their layer sizes are $37$--$256$--$256$ (element encoder), $269$--$256$--$256$ (context encoder, on the mean element embedding and the $13$ context features), $512$--$256$--$5$ (element head, whose five outputs parameterize $\beta_i^{\rm R}$, $\Delta\varphi_i$, and $\Delta\vartheta_i$), and $302$--$256$--$256$--$33$ (beamformer head, whose $33$ outputs parameterize the power split, the coefficients over $\mathcal{B}$, and $\bs{b}$ in \eqref{eq:decoder}), which gives $4.95\times10^{5}$ trainable parameters.
These sizes do not depend on $M$ or $N$, since the element networks are shared by all elements and the beamformer head acts on the basis \eqref{eq:basis}.
For a new size, only the weights and the feature normalizer are refitted.

\subsection{Training With an Analytic Critic}\label{sec:training}
Deterministic actor--critic methods \cite{silver2014deterministic} update the actor parameters $\bs{\theta}$ along the policy gradient
\begin{equation}
\nabla_{\bs{\theta}}J=\Exp_{\bs{o}}\bigl\{\nabla_{\bs{\theta}}\pi_{\bs{\theta}}(\bs{o})\,\nabla_{\bs{\xi}}Q^{\pi}(\bs{o},\bs{\xi})\big|_{\bs{\xi}=\pi_{\bs{\theta}}(\bs{o})}\bigr\},
\label{eq:dpg}
\end{equation}
where $J$ is the expected reward and the action-value function $Q^{\pi}$ is the expected immediate reward in the single-step setting of Section~\ref{sec:mdp}.
Model-free learners estimate this gradient from sampled rewards \cite{fujimoto2018addressing, haarnoja2018soft, schulman2017proximal}.
In contrast, the covert rate and the constraint rates in \eqref{eq:qos} are known differentiable functions of the action and channel.
Their exact action gradients follow by differentiating through the system model, which acts as an analytic critic as in model-based value-gradient methods \cite{heess2015learning}.
Hence, no value function, exploration noise, or solver labels are needed, and the update coincides with that of unsupervised learning-to-optimize methods \cite{xia2020deep, gao2020unsupervised}.
Since the feasibility indicator in $S$ has zero gradient almost everywhere, the actor minimizes the surrogate loss
\begin{equation}
\mathcal{L}(\bs{\theta})=-\Exp\bigl\{\log R_{\rm f}^{\rm cov}+\log\Lambda_{\omega}\bigr\},
\label{eq:logloss}
\end{equation}
where $\Lambda_{\omega}$ is the product of $\sig(\omega(x/x^{\rm th}-1))$ over the rate constraints $x\ge x^{\rm th}$ in \eqref{eq:qos}, and $\omega$ sets the sharpness.
As $\omega$ increases, $\Lambda_{\omega}$ approaches the feasibility indicator.
The actor gradient is \eqref{eq:dpg} with $Q^{\pi}$ replaced by the negative per-sample loss.
For a strongly violated constraint with argument $z\ll0$, $\log\sig(z)\approx z$, which keeps the gradient nonzero in the infeasible region.
The loss \eqref{eq:logloss} has no covertness penalty, since the decoder \eqref{eq:decoder} guarantees covertness.
Algorithm~\ref{alg:train} summarizes the training.

\begin{algorithm}[!t]
\caption{Training of the Proposed Actor}
\label{alg:train}
\begin{algorithmic}[1]
\Require training set $\mathcal{S}_{\rm tr}$, development set $\mathcal{S}_{\rm dev}$, $T_{\rm s}$ steps, batch size $N_{\rm b}$
\State fit the feature normalizer on $\mathcal{S}_{\rm tr}$ ($\bs{R}_{h}$ as in Section~\ref{sec:res-setup})
\For{$t=0,\ldots,T_{\rm s}-1$}
  \State decrease the step size, and increase $\omega$ if at least $90\,\%$ of the designs of the last ten batches are feasible
  \State sample $N_{\rm b}$ channels and compute $\bs{\xi}=\pi_{\bs{\theta}}(\bs{o}(\bs{H}))$ with the decoder \eqref{eq:decoder}
  \State take an Adam step on $\nabla_{\bs{\theta}}\mathcal{L}$ of \eqref{eq:logloss} through the system model
\EndFor
\Ensure actor $\pi_{\bs{\theta}}$ after the last step, selected as the best of three independent training runs on $\mathcal{S}_{\rm dev}$
\end{algorithmic}
\end{algorithm}

\subsection{Complexity Analysis}\label{sec:complexity}
One forward pass of the actor requires $\mathcal{O}(N(FD+D^{2})+MN^{2}+M^{3})$ floating-point operations (FLOPs), with $F=37$ features per element and $D=256$ hidden neurons per layer, where the $M^{3}$ term comes from the whitening in \eqref{eq:basis}.
The total is $6.3\times10^{7}$ FLOPs for $M=32$ and $N=128$.
One training run of Algorithm~\ref{alg:train} costs $1.0\times10^{14}$ FLOPs, equal to the per-channel solver's gradient steps on about $1.6\times10^{3}$ channel blocks.

\section{Numerical Results and Discussion}\label{sec:results}

\subsection{Simulation Settings and Benchmarks}\label{sec:res-setup}

\begin{table}[!t]
\centering
\caption{Simulation Parameters}
\label{tab:params}
\renewcommand{\arraystretch}{1.0}\footnotesize
\setlength{\tabcolsep}{2.5pt}
\begin{tabular}{ll}
\hline
\textbf{Parameter} & \textbf{Value (range)} \\
\hline
Number of BS antennas $M$ & 32 (16--128) \\
Number of STAR-RIS elements $N$ & $128=8\times16$ (64--256) \\
Carrier frequency & 3.5 GHz \\
Bandwidth & 20 MHz \\
Noise figure & 7 dB \\
Path loss per hop & $(\lambda_{\rm c}/4\pi)^{2}d^{-\eta}$ with $\eta=2$ \\
BS angular spread $\Delta$ & $15^{\circ}$ \\
BS--STAR-RIS distance & 50 m \\
STAR-RIS--node distances $d_{\rm n}$, $d_{\rm f}$, $d_{\rm w}$ & 4 m, 6 m, 6 m \\
Transmit power budget $P_{S}$ & 23 dBm (20--40 dBm) \\
Number of warden samples $n$ & 15 (15--$10^{3}$) \\
Covertness budget $\epsilon$ & 0.1 (0.05--0.2) \\
Public requirement $R_{\rm th}$ & 5 bps/Hz (2--8 bps/Hz) \\
Common-rate floor $R_{\rm c}^{\rm th}$ & 0.5 bps/Hz \\
Covert-rate floor $R_{\rm f}^{\rm th}$ & 0.6 bps/Hz \\
\hline
\end{tabular}
\end{table}

\begin{table*}[!t]
\centering
\caption{Covert Rate (bps/Hz) on 1024 Unseen Channels, CPU Time per Design on One Thread, and Training FLOPs}
\label{tab:main}
\renewcommand{\arraystretch}{1.0}\footnotesize\setlength{\tabcolsep}{3.2pt}
\begin{tabular*}{\textwidth}{@{\extracolsep{\fill}}lcccccccccccc@{}}
\hline
\multirow{2}{*}{Scheme} & \multicolumn{6}{c}{$R_{\rm th}=5$~bps/Hz, $P_S$ (dBm)} & \multicolumn{4}{c}{\makebox[0pt]{$P_S=23$~dBm, $R_{\rm th}$ (bps/Hz)}} & CPU time & Training \\
 & 20 & 23 & 26 & 30 & 35 & 40 & 2 & 3 & 4 & 6 & per design & FLOPs/run \\
\hline
\multicolumn{13}{l}{\emph{Non-learned designs}}\\
Fast closed forms, RSMA or NOMA & 4.22 & 5.28 & 6.44 & 7.89 & 9.60 & 11.30 & 6.71 & 6.71 & 6.43 & 4.44 & 23 ms & none \\
Best closed form, RSMA or NOMA & 4.23 & 5.28 & 6.44 & 7.89 & 9.61 & 11.30 & 6.71 & 6.71 & 6.43 & 4.45 & 55 s & none \\
NOMA closed form & 4.22 & 5.24 & 6.42 & 7.88 & 9.60 & 11.27 & 6.71 & 6.71 & 6.43 & 4.22 & 17 ms & none \\
AO+CEM & 0.00 & 0.05 & 0.40 & 1.71 & 3.45 & 4.80 & 2.53 & 1.99 & 0.63 & 0.00 & 1.8 s & none \\
Per-channel RSMA solver (offline reference) & 4.96 & 6.10 & 7.21 & 8.65 & 10.40 & 12.10 & 6.71 & 6.70 & 6.39 & 5.73 & 60 s & none \\
Average-DEP solver (offline, not certified) & 3.43 & 4.20 & 5.20 & 6.85 & 9.39 & 12.02 & 6.39 & 6.31 & 4.56 & 4.09 & $>$60 s & none \\
\hline
\multicolumn{13}{l}{\emph{Proposed certified DRL}}\\
\textbf{Single actor, all ten points} & 4.35 & 5.52 & 6.64 & 8.10 & 9.84 & 11.52 & 6.21 & 6.08 & 5.84 & 5.12 & 3.3 ms & $1.0\times10^{15}$ \\
\textbf{Per-point actor} & 4.26 & 5.43 & 6.51 & 7.96 & 9.59 & 11.15 & 6.23 & 5.99 & 5.71 & 5.03 & 3.3 ms & $1.0\times10^{14}$ \\
\textbf{Per-point actor + 256 certified steps} & 4.80 & 5.93 & 6.97 & 8.34 & 9.80 & 11.27 & 6.50 & 6.31 & 6.11 & 5.58 & 0.64 s & $1.0\times10^{14}$ \\
\hline
\multicolumn{13}{l}{\emph{Same decoder and analytic critic}}\\
Linear policy & 4.26 & 5.36 & 6.39 & 7.81 & 9.23 & 10.75 & 6.21 & 5.94 & 5.64 & 4.96 & 3.2 ms & $1.0\times10^{14}$ \\
Conventional RIS & 2.84 & 3.89 & 4.89 & 6.24 & 7.89 & 9.53 & 4.49 & 4.13 & 4.11 & 3.51 & 3.3 ms & $1.0\times10^{14}$ \\
\hline
\multicolumn{13}{l}{\emph{Model-free DRL, same decoder}}\\
PPO, tuned & 1.53 & 2.54 & 3.75 & 4.92 & 6.18 & 7.13 & 5.32 & 3.64 & 2.63 & 1.85 & 3.3 ms & $4.0\times10^{14}$ \\
PPO, untuned & 0.03 & 0.00 & 3.54 & 5.19 & 0.00 & 8.23 & 2.28 & 1.94 & 0.00 & 0.07 & 3.3 ms & $4.0\times10^{14}$ \\
SAC & 0.00 & 0.71 & 0.52 & 1.11 & 2.41 & 3.61 & 4.38 & 2.02 & 0.57 & 0.12 & 3.3 ms & $1.5\times10^{15}$ \\
TD3 & 0.77 & 1.55 & 1.37 & 1.51 & 2.16 & 3.13 & 3.70 & 1.83 & 1.18 & 0.00 & 3.3 ms & $7.8\times10^{14}$ \\
SAC, full action & 0.00 & 0.00 & 0.00 & 0.00 & 0.00 & 0.00 & 3.58 & 0.66 & 0.00 & 0.00 & 3.4 ms & $1.8\times10^{15}$ \\
TD3, full action & 0.00 & 0.00 & 0.00 & 0.00 & 0.00 & 0.00 & 0.00 & 0.00 & 0.00 & 0.00 & 3.3 ms & $9.9\times10^{14}$ \\
\hline
\multicolumn{13}{l}{\emph{Discrete-action RL over 420 certified designs}}\\
Tabular bandit & 1.53 & 2.32 & 3.20 & 4.46 & 6.08 & 7.74 & 5.21 & 4.32 & 2.35 & 1.94 & 2.3 ms & $9.3\times10^{12}$ \\
DQN & 1.50 & 2.30 & 3.17 & 4.44 & 6.08 & 7.73 & 5.17 & 4.26 & 2.29 & 1.87 & 2.4 ms & $1.2\times10^{13}$ \\
\hline
\end{tabular*}
\end{table*}

This section validates the analysis and evaluates the proposed framework by Monte Carlo simulations.
Unless otherwise stated, the parameters of Table~\ref{tab:params} are used, with the number of warden samples of \cite{vu2025performance} and the rate floors of \cite{tu2025active}.
Finite-blocklength effects are not considered.
Since a certified design depends on $(n,\epsilon)$ only through $\varrho$, changing the budget $\epsilon$ is equivalent to changing $n$.
The operating points above the crossover \eqref{eq:crossover}, i.e., $R_{\rm th}>4.94$~bps/Hz, are referred to as the \emph{demanding regime}.
For $\bs{R}_{h}$, the BS uses the sample covariance of the STAR-RIS--warden channel over draws with the warden's angle perturbed within $\pm2^{\circ}$, shrunk toward the scaled identity, which equals $d_{\rm w}^{-\eta}\bs{I}_{N}$ up to sampling error.

We evaluate three variants of the proposed framework, namely a per-point actor (one per operating point), a single actor trained for all operating points, and the per-point actor refined by $256$ certified gradient steps.
The actors are trained by Algorithm~\ref{alg:train} with $T_{\rm s}=2000$ steps, $N_{\rm b}=256$, $\omega$ annealed from $20$ to $200$, a cosine step size from $3\times10^{-4}$ to $3\times10^{-6}$, and gradient clipping at $1$.
The single actor runs $20000$ steps, cycling its batches over the ten operating points.
For a fair comparison, all learners share the same decoder, reward, training channels, and number of training draws per operating point.
The benchmark schemes are as follows.
\begin{enumerate}
\item The closed forms of Section~\ref{sec:cf} search $\beta^{\rm R}$, $\tau$, the shield, and the private fraction or margin on scalar grids of $10^{2}$ to $3\times10^{4}$ points per channel.
The fast closed forms replace the grid searches by bisection.
\item A per-channel certified RSMA solver serves as an offline reference.
It runs $2000$ Adam steps of size $0.1$ from the best RSMA closed form and from a singular-value start, keeps the best feasible iterate of each run, and reports the per-channel maximum of the two runs and the RSMA closed forms.
It cannot represent the NOMA design ($\tau=0$) and can fall below the NOMA closed form.
\item AO+CEM is AO with a cross-entropy method (CEM) that imposes covertness on the believed powers \eqref{eq:power}.
It starts from matched-filter precoders and the aligned phases and uses $8000$ objective evaluations per channel.
The average-DEP solver is the solver above with the certificate replaced by the average-DEP criterion $\Exp_{\bs{h}_{\rm w}}\{\mathrm{DEP}^{\star}\}\ge1-\epsilon$ under $\bs{R}_{h}$, in line with the average-type constraints of \cite{wu2021intelligent, xiao2024star, nguyen2025enhancing}.
It runs $2000$ Adam steps of size $0.03$ from the certified solver's design with free precoders, estimates the average DEP from $1024$ draws, and transmits at the largest power meeting the constraint on fresh draws, which is $P_{S}$ on all channels.
The optimizers of \cite{xiao2024star, nguyen2025enhancing} are not reproduced.
\item In model-free DRL, PPO \cite{schulman2017proximal} acts on the full action through a Gaussian policy around the output of the proposed actor, with a clipped surrogate (clip $0.2$, four passes per batch, and an entropy bonus).
It is trained with default hyperparameters and with the step size tuned on the development channels over $\{10^{-4},3\times10^{-4},10^{-3}\}$ and a constant or cosine schedule, which selects $10^{-4}$.
Soft actor--critic (SAC) \cite{haarnoja2018soft} and TD3 \cite{fujimoto2018addressing} output the coefficients over \eqref{eq:basis} and the powers on a surface fixed at the aligned phases \eqref{eq:align} with $\beta_i^{\rm R}=1/2$.
Their twin layer-normalized critics observe the $13$ context features and the action and regress the logarithmic reward (horizon one) with $16$ updates per step.
To separate the effect of the critic from that of the action space, SAC and TD3 are also trained on the actor's full action, using the proposed actor network and critics that observe its full normalized state.
For these variants, three settings each of the TD3 exploration noise and the SAC initial temperature were tried on the development channels at $23$~dBm and $R_{\rm th}=5$~bps/Hz.
All yield a zero covert rate, so the smallest is kept.
\item In discrete-action RL, a tabular bandit and a DQN \cite{mnih2015human} select among $420$ certified designs.
Each design combines one of $84$ colinear designs on the fixed surface of SAC and TD3 (matched or whitened filters for each user, three power gaps, and seven shield levels) with one of five values of $\tau$.
The DQN uses two hidden layers of $256$ units, random-action exploration, and a $10^{5}$-sample replay memory.
\item As ablations with the same decoder and critic, we train a linear policy, i.e., the actor with identity activations, and the actor on a conventional RIS with half reflect-only and half transmit-only elements, both by Algorithm~\ref{alg:train} with the above settings.
\end{enumerate}
All learners are trained on the same $4096$ channels in three independent runs, and the reported run is selected on $128$ development channels.
All schemes are evaluated on the same $1024$ unseen channels, disjoint from the training and development sets, and all learners at their final iterates.
All schemes and the reported training runs were fixed before any evaluation on the unseen channels, where the bandit and DQN runs were selected on an earlier disjoint channel set.
The covert rate of a design that violates a rate constraint or is detected, i.e., violates \eqref{eq:covertness} on the realized warden channel, is counted as zero.
A zero entry in Table~\ref{tab:main} thus means that no feasible covert design was found on any unseen channel.
Confidence intervals of the paired differences are percentile bootstrap intervals with $10^{4}$ resamples, and those of the main comparisons are Bonferroni-corrected to hold simultaneously over their operating points.
The $95\,\%$ confidence half-widths of the entries of Table~\ref{tab:main} do not exceed $0.07$~bps/Hz, except for AO+CEM and the average-DEP solver ($0.24$~bps/Hz).
CPU times are measured in double precision (PyTorch, NumPy) on one thread of an AMD Ryzen Threadripper 7960X at $P_S=23$~dBm and $R_{\rm th}=5$~bps/Hz.
All schemes are timed with batch size one on one pinned thread, as in Table~\ref{tab:latency}, except the best closed form, AO+CEM, and the per-channel solver, which use their original implementations.
The learner rows of Table~\ref{tab:main} were timed in a separate session, in which the actors took $7$--$9\,\%$ less time than in that of Table~\ref{tab:latency}.

\subsection{Results and Discussion}\label{sec:res-disc}

\subsubsection{Convergence Behavior}\label{sec:res-conv}
\begin{figure}[!t]
\centering
\includegraphics[width=\columnwidth]{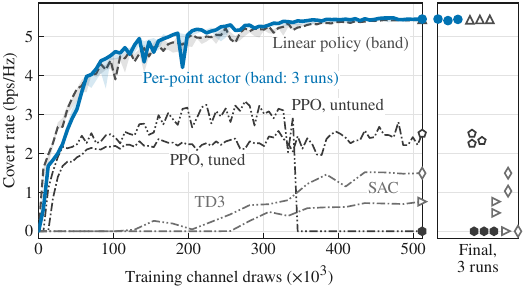}
\caption{Convergence behavior of the learners, where $P_{S}=23$~dBm and $R_{\rm th}=5$~bps/Hz.}
\label{fig:learning}
\end{figure}

Fig.~\ref{fig:learning} illustrates the convergence behavior of the learners on the development channels, where the bands show the range of the three runs and the right panel the final value of every run.
The per-point actor converges smoothly within the training budget, and the linear policy with the same critic follows the same curve.
In contrast, PPO with the default step size collapses to infeasible designs in at least two of three training runs at nine of the ten operating points, which explains its zero entries in Table~\ref{tab:main}.
Tuned PPO stays well below the actor, and SAC and TD3 learn slowly.
This gap is mainly due to the analytic critic, which returns the exact constraint gradients, whereas PPO estimates them from likelihood ratios and SAC and TD3 fit a critic of the shaped reward from samples.
Since SAC and TD3 act on a fixed surface, their gap also reflects the smaller action space.
Even with the full action and full-state critics, neither SAC nor TD3 finds a design that meets the public requirement in the demanding regime in any training run (Table~\ref{tab:main}).
The deterministic full-action TD3 actor saturates at its action bound within $100$ updates in every training run.
Across the three training runs, the mean covert rates of the proposed actors at $R_{\rm th}=5$~bps/Hz differ by at most $1.7\,\%$, whereas those of tuned PPO, SAC, and TD3 differ by up to $2.6$, $0.7$, and $1.5$~bps/Hz, respectively, and those of untuned PPO range from $0$ to $8.2$~bps/Hz at $40$~dBm.

\subsubsection{Certified Covertness}\label{sec:res-cov}
\begin{figure}[!t]
\centering
\includegraphics[width=\columnwidth]{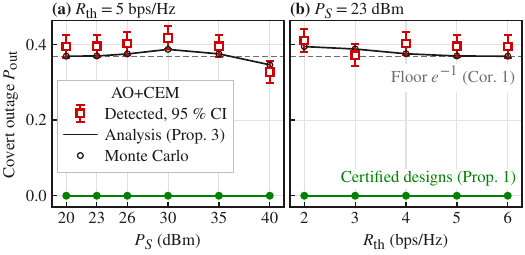}
\caption{Covert outage probability of AO+CEM and of the certified designs versus (a) the transmit power, where $R_{\rm th}=5$~bps/Hz, and (b) the public requirement, where $P_{S}=23$~dBm.}
\label{fig:outage}
\end{figure}

Fig.~\ref{fig:outage} shows the covert outage probability of AO+CEM, which constrains only the believed powers, with Wilson $95\,\%$ intervals over $1024$ channels and the floor $e^{-1}$ of Corollary~\ref{cor:floor} (dashed).
As predicted by Propositions~\ref{prop:shield} and~\ref{prop:multi} and Corollary~\ref{cor:hybrid}, none of the $5.4\times10^{5}$ certified designs evaluated on the unseen channels is detected on its realized warden channel, which gives their zero outage in Fig.~\ref{fig:outage}.
The same holds for the hybrid realizations with two, three, and four phase-shifter RF chains, which are evaluated at five operating points.
With four chains, both actors' precoders reproduce the digital design within a relative error of $1.6\times10^{-14}$, and with three chains, the per-point actor and the single actor retain $95.6$--$98.3\,\%$ and $80.5$--$99.7\,\%$ of the covert rate, respectively, while at most $9\,\%$ remains with two chains.
In contrast, the fraction of unseen channels on which AO+CEM is detected is close to the expression of Proposition~\ref{prop:outage} and to the Monte Carlo estimate ($2\times10^{4}$ warden channels per design), which agree at all operating points, and thus close to or above the floor $e^{-1}$.
Its power cap \eqref{eq:power} also leaves the public requirement unmet on most channels below $30$~dBm at $R_{\rm th}=5$~bps/Hz and at $6$~bps/Hz, which explains its low covert rate in Table~\ref{tab:main}.
The average-DEP solver meets its own criterion, with a mean realized DEP of $0.91$--$0.92$.
On the channels on which it is detected, the median realized DEP is $0.88$--$0.89$.
At $20$--$26$~dBm and at $R_{\rm th}=4$ and $6$~bps/Hz, it gains $4$--$15\,\%$ over the certified solver by its own criterion but is detected on $31$--$40\,\%$ of the unseen channels, whereas at the other operating points it gains less than $2\,\%$ and is detected on $0.8$--$22\,\%$.
Counted as in Table~\ref{tab:main}, it stays below the certified solver at all operating points.

Finally, we consider a warden outside the threat model of Assumption~\ref{ass:warden}, which is located at the same distance and decodes and perfectly cancels the cover.
Since a low-rate cover is easier to decode, the learned designs of Table~\ref{tab:main} are exposed on $58$--$91\,\%$ of the channels at $R_{\rm th}=5$~bps/Hz from $20$ to $30$~dBm and on $64$--$77\,\%$ at $6$~bps/Hz.
In contrast, the NOMA and colinear closed forms, whose cover carries the public message, remain covert on all unseen channels up to $30$~dBm.

\subsubsection{Performance Comparison}\label{sec:res-regime}
In Table~\ref{tab:main}, up to $R_{\rm th}=4$~bps/Hz the closed forms match the per-channel solver, with the NOMA closed form being the best design at $4$~bps/Hz.
From $R_{\rm th}=5$~bps/Hz onward, the per-channel RSMA solver outperforms the closed forms at all transmit powers, with a gap that narrows as the power increases.
As anticipated in Section~\ref{sec:mdp}, the solver serves the public user through a zero-forced private stream, whereas the colinear and NOMA closed forms use the common stream, which reduces their admissible shield as $R_{\rm th}$ grows (Proposition~\ref{prop:rsma}(\textit{ii})).
The crossover between $4$ and $5$~bps/Hz agrees in location with the high-SNR prediction of Remark~\ref{rem:crossover}, although the finite-power gaps are smaller.

In the demanding regime, the single actor outperforms the best closed form at all transmit powers, with simultaneous confidence intervals above zero (lower bounds of at least $0.08$~bps/Hz).
The gain grows with the public requirement and reaches $15.2\,\%$ at $R_{\rm th}=6$~bps/Hz.
This holds for each of its three training runs, while the per-point actors match the best closed form at $35$~dBm and fall $1.3\,\%$ below it at $40$~dBm.
The single actor uses the same total training FLOPs as the ten per-point actors (Table~\ref{tab:main}) but receives ten times as many updates as each of them, which may explain this difference.
Below the crossover, the closed forms, which already match the solver, outperform the learned designs.

\begin{table}[!t]
\centering
\caption{Covert Rate (bps/Hz) Versus the Covertness Budget $\epsilon$ on 1024 Unseen Channels}
\label{tab:eps}
\renewcommand{\arraystretch}{1.0}\footnotesize\setlength{\tabcolsep}{2pt}
\begin{tabular*}{\columnwidth}{@{\extracolsep{\fill}}lccc@{}}
\hline
Covertness budget $\epsilon$ & 0.05 & 0.10 & 0.20 \\
$\varrho$ & 0.033 & 0.067 & 0.141 \\
Crossover \eqref{eq:crossover} (bps/Hz) & 5.94 & 4.94 & 3.93 \\
\hline
Best closed form & 5.24 & 5.28 & 5.71 \\
Per-channel RSMA solver & 5.32 & 6.10 & 6.88 \\
Single actor & 4.74 & 5.52 & 6.28 \\
Per-point actor & 4.59 & 5.43 & 6.22 \\
Single actor / best closed form & 0.903 & 1.046 & 1.098 \\
\quad paired $95\,\%$ interval & [0.899, 0.907] & [1.041, 1.051] & [1.091, 1.105] \\
\hline
\end{tabular*}
\end{table}

\subsubsection{Impact of the Covertness Budget}\label{sec:res-eps}
Table~\ref{tab:eps} shows the impact of the covertness budget $\epsilon$ at $P_{S}=23$~dBm and $R_{\rm th}=5$~bps/Hz, where changing $\epsilon$ moves the crossover \eqref{eq:crossover} across this operating point.
At $\epsilon=0.05$, the crossover rises above the operating point, the NOMA closed form is the best closed form on all channels, and the actors trained for this budget fall $10$--$12\,\%$ below it, since the NOMA design ($\tau=0$) lies outside the actor's design set.
At $\epsilon=0.2$, the crossover falls to $3.93$~bps/Hz and the gap between the solver and the best closed form widens.
Accordingly, the single actor's gain over the best closed form increases from $4.6\,\%$ at $\epsilon=0.1$ to $9.8\,\%$ at $\epsilon=0.2$.

\begin{figure}[!t]
\centering
\includegraphics[width=\columnwidth]{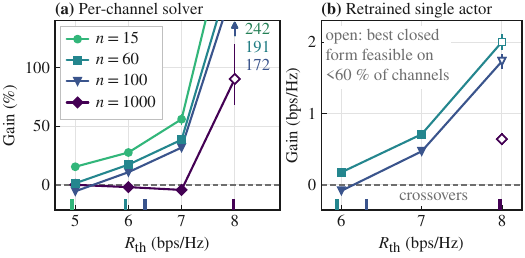}
\caption{Gain over the best closed form versus $R_{\rm th}$ for (a) the per-channel solver and (b) the retrained single actor, where $P_{S}=23$~dBm.}
\label{fig:nsamples}
\end{figure}

\subsubsection{Impact of the Warden's Sample Count}\label{sec:res-n}
Fig.~\ref{fig:nsamples}(a) shows the gap between the per-channel solver and the best closed form on the development channels for each $n$, with the crossover \eqref{eq:crossover} of each $n$ marked and the values above $100\,\%$ printed.
The gap appears at most $1$~bps/Hz below each predicted crossover and widens above it.
Hence, as $n$ grows, the regime in which learning can help shifts to stricter public requirements, consistent with Remark~\ref{rem:crossover}.
Fig.~\ref{fig:nsamples}(b) shows the gain of the single actor retrained for each $n$ on $512$ separate unseen channels at the operating points with a gap.
Without retraining, the actor trained for $n=15$ returns no feasible design at $n\ge60$.
The retrained actor outperforms the best closed form at all of these points except at $n=100$ and $R_{\rm th}=6$~bps/Hz, a point below its crossover, where it is slightly worse.
At $R_{\rm th}=8$~bps/Hz, the best closed form meets the public requirement on fewer than $60\,\%$ of the channels, so that the gain comes mainly from channels on which only the actor meets it.

\begin{table}[!t]
\centering
\caption{Covert Rate (bps/Hz) Versus $M$ and $N$ on 512 Unseen Channels}
\label{tab:size}
\renewcommand{\arraystretch}{1.0}\footnotesize\setlength{\tabcolsep}{2.2pt}
\begin{tabular*}{\columnwidth}{@{\extracolsep{\fill}}lcccccc@{}}
\hline
\multirow{2}{*}{Scheme} & \multicolumn{4}{c}{$N=128$, $M$} & \multicolumn{2}{c}{$M=32$, $N$} \\
 & 16 & 32 & 64 & 128 & 64 & 256 \\
\hline
Best closed form & 4.54 & 5.28 & 6.10 & 6.97 & 3.53 & 7.26 \\
Per-channel solver & 5.26 & 6.10 & 6.89 & 7.74 & 4.10 & 8.02 \\
Single actor & 4.68 & 5.50 & 6.28 & 7.16 & 3.39 & 7.47 \\
Gain (\%) & 3.2 & 4.3 & 3.0 & 2.8 & $-$3.9 & 2.9 \\
\hline
\end{tabular*}
\end{table}

\subsubsection{Impact of the Array and Surface Sizes}\label{sec:res-size}
Table~\ref{tab:size} shows the covert rate versus $M$ and $N$ at $P_S=23$~dBm and $R_{\rm th}=5$~bps/Hz, where the single actor is trained for each size.
The gap between the per-channel solver and the best closed form narrows slowly as $M$ or $N$ grows, since the array gain raises the effective SNR, but exceeds $10\,\%$ at all sizes.
The single actor outperforms the best closed form at all sizes except $N=64$, where it falls below it in all three training runs.
This is mainly because the actor stays $0.55$--$0.71$~bps/Hz below the solver at every size, whereas the solver's room over the best closed form is only $0.57$~bps/Hz at $N=64$ and at least $0.72$~bps/Hz at the other sizes.
Without retraining, it returns no feasible design at a new size.

\subsubsection{Learning Benchmarks and Ablations}\label{sec:res-abl}
With the same decoder, the per-point actor outperforms the best of the eight model-free and discrete-action learners in Table~\ref{tab:main} by more than one third at all demanding operating points.
It achieves $1.36$--$2.78$ times the covert rate of the best other learner, with simultaneous lower bounds of at least $1.34$.
DQN performs no better than the tabular bandit over the same discrete action set, and over the codebook of \cite{wang2022intelligent} the discrete-action learners find no feasible design above $R_{\rm th}=2$~bps/Hz.
The conventional RIS falls below the STAR-RIS at all operating points, which quantifies the per-element energy split.
The linear policy matches the deep actor at $20$~dBm and stays slightly below it elsewhere.
It remains above the best closed form at $20$ and $23$~dBm and at $6$~bps/Hz, but not from $26$~dBm upwards.
Hence, the gain over the model-free learners stems mainly from the exact gradient, with a small additional gain from depth at high transmit power.

\begin{table}[!t]
\centering
\caption{Covert Rate (bps/Hz) Under the Same Single-Thread Computation Time on 256 Additional Unseen Channels}
\label{tab:latency}
\renewcommand{\arraystretch}{1.0}\footnotesize\setlength{\tabcolsep}{2.2pt}
\begin{tabular*}{\columnwidth}{@{\extracolsep{\fill}}lccccccc@{}}
\hline
\multirow{2}{*}{Scheme} & \multicolumn{6}{c}{$R_{\rm th}=5$~bps/Hz, $P_S$ (dBm)} & $R_{\rm th}=6$ \\
 & 20 & 23 & 26 & 30 & 35 & 40 & 23 dBm \\
\hline
\multicolumn{8}{l}{\emph{At the actor's time ($3.4$--$3.8$~ms)}}\\
Single actor & 4.34 & 5.52 & 6.67 & 8.12 & 9.87 & 11.56 & 5.12 \\
Colinear closed form & 3.77 & 4.84 & 5.87 & 7.37 & 9.11 & 10.80 & 3.76 \\
Gain (\%) & 15.0 & 14.0 & 13.6 & 10.2 & 8.3 & 7.0 & 36.0 \\
\hline
\multicolumn{8}{l}{\emph{Within the coherence time $T_{\rm c}=43.5$~ms}}\\
Single actor (+ steps) & 4.35 & 5.52 & 6.67 & 8.12 & 9.87 & 11.56 & 5.13 \\
Best non-learned & 4.26 & 5.30 & 6.46 & 7.91 & 9.64 & 11.33 & 4.47 \\
Gain (\%) & 2.0 & 4.1 & 3.2 & 2.7 & 2.4 & 2.0 & 14.7 \\
\hline
\end{tabular*}
\end{table}

\begin{figure}[!t]
\centering
\includegraphics[width=\columnwidth]{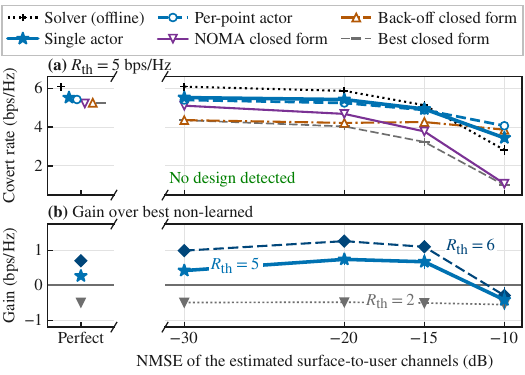}
\caption{(a) Covert rate and (b) gain of the single actor over the best non-learned design versus the NMSE of the estimated surface-to-user channels, where $P_{S}=23$~dBm.}
\label{fig:csi}
\end{figure}

\subsubsection{Imperfect Legitimate CSI}\label{sec:res-csi}
Fig.~\ref{fig:csi} shows the impact of imperfect CSI of the STAR-RIS--user channels, where all schemes compute their designs from the estimates $\hat{\bs{h}}_{k}=\bs{h}_{k}+\bs{e}_{k}$, with $\bs{e}_{k}\sim\mathcal{CN}(\bs{0},s^{2}d_{k}^{-\eta}\bs{I}_{N})$ and $s^{2}$ the normalized mean-square error (NMSE).
The designs are evaluated on $512$ unseen channels against the true channels, without retraining or retuning.
Consistent with Propositions~\ref{prop:shield} and~\ref{prop:multi}, no design is detected on the realized warden channel at any NMSE.
The covert rate of the closed forms decreases faster with the NMSE than that of the learned designs, which retain most of it up to $-15$~dB.
This is partly because the closed forms meet the public requirement with almost no margin, so that an estimation error violates it on the true channel, whereas the learned designs, which do not model CSI errors, leave a small margin.
With perfect CSI, the median public-rate margin above $R_{\rm th}=5$~bps/Hz is $0.008$~bps/Hz for the best closed form and $0.21$~bps/Hz for the single actor.
We also evaluate a back-off closed form, i.e., the fast closed forms designed for $R_{\rm th}+\delta$ on the estimate, where $\delta\le0.5$~bps/Hz is selected a posteriori per NMSE level.
It improves on the NOMA closed form only at $-15$ and $-10$~dB for $R_{\rm th}=5$ and $6$~bps/Hz.
Against the stronger of these two designs at each NMSE (Fig.~\ref{fig:csi}(b)), from $-30$ to $-15$~dB, the single actor remains $0.4$--$0.7$ and $1.0$--$1.3$~bps/Hz higher at $R_{\rm th}=5$ and $6$~bps/Hz, respectively, and falls below it at $-10$~dB and at $R_{\rm th}=2$~bps/Hz.

\subsubsection{Refinement and Execution Time}\label{sec:res-refine}
With $256$ certified gradient steps, the refined per-point actor approaches the covert rate of the per-channel solver with about two orders of magnitude less computation time (Table~\ref{tab:main}).
In the demanding regime, it outperforms the best closed form except at $40$~dBm.
With the same number of steps, the learned initialization yields a higher covert rate than a closed-form or singular-value start at $23$--$30$~dBm and at $R_{\rm th}=6$~bps/Hz, and a comparable one at $20$~dBm.

Table~\ref{tab:latency} compares the single actor with the non-learned certified designs under the same computation time on one CPU thread, where each component uses its fastest implementation that yields the same designs.\footnote{The reported times are specific to this machine and a single CPU thread. Graphics processing units and batched inference are not considered.}
The coherence time is $T_{\rm c}=0.423\lambda_{\rm c}/v=43.5$~ms at a speed of $v=3$~km/h \cite{rappaport2002wireless}, each certified step counts $2.45$~ms, and the colinear closed form needs $2.4$--$3.1$~ms.
Within $T_{\rm c}$, the single actor, alone or followed by the certified steps that fit in the remaining time, outperforms the best non-learned certified design computable within $T_{\rm c}$ at all demanding operating points.
This gain has paired confidence intervals above zero and holds for all three training runs.
The gain increases with the public requirement and is larger at the execution time of the actor, where only the colinear closed form can be computed.
This is because one forward pass of the actor costs $6.3\times10^{7}$ FLOPs (Section~\ref{sec:complexity}), whereas the fast closed forms build and evaluate about $40$ candidate designs by bisection ($0.37$~ms each, $12$--$30$~ms in total, or $5.4$--$8.4$~ms without the NOMA closed form) and the solver runs thousands of gradient steps per channel (about a minute).

\section{Concluding Remarks and Open Challenges}\label{sec:conclusion}
In this work, we have studied covert rate maximization in a STAR-RIS-assisted RSMA massive MIMO downlink with all direct links blocked and an unknown warden channel.
Confining the covert precoder to the span of the cover precoders with a bounded norm certifies covertness against an energy-detecting warden for every warden channel.
On this basis, we derived certified closed forms and trained a DRL actor with the differentiable system model as an analytic critic.
Numerical results showed that the closed forms match the per-channel solver at mild public requirements and that the crossover moves with the covertness budget as predicted by \eqref{eq:crossover}.
Above the crossover, the single actor outperforms all non-learned certified designs under the same single-thread computation time within a coherence block, by up to $14.7\,\%$, as well as model-free DRL and classical RL over certified designs.

Several directions remain open.
First, a warden that decodes and cancels the cover \cite{ahmadi2026square, wang2026joint} requires new certificates, since it exposes the learned designs on most channels (Section~\ref{sec:res-cov}).
Second, estimated and outdated CSI would turn the design into a multi-step decision process.
Finally, an active surface with amplification noise and a reflect-power constraint \cite{zhang2023active} and a varying number of users remain to be studied.

\appendices

\section{Proofs of Propositions~\ref{prop:shield} and~\ref{prop:multi}}\label{app:shield}
\emph{Proposition~\ref{prop:shield}.} (\textit{i}) If $\bs{w}_{\rm f}=\bs{W}\bs{b}$, the Cauchy--Schwarz inequality gives $|\bs{a}^{H}\bs{W}\bs{b}|^{2}\le\|\bs{W}^{H}\bs{a}\|^{2}\|\bs{b}\|^{2}$, so that $\rho(\bs{a})\le1+\|\bs{b}\|^{2}P\|\bs{W}^{H}\bs{a}\|^{2}/(P\|\bs{W}^{H}\bs{a}\|^{2}+\sigma_{\rm w}^{2})\le1+\|\bs{b}\|^{2}\le\rho^{\star}$ for every $\bs{a}$, $P$, and $\sigma_{\rm w}^{2}>0$.
The noise enters only the denominator, so noise that depends on the channel can only lower $\rho$.
(\textit{ii}) If $\bs{w}_{\rm f}\notin\operatorname{span}(\bs{W})$, the component of $\bs{w}_{\rm f}$ orthogonal to $\operatorname{span}(\bs{W})$ lies in $\mathcal{S}$.
Taking it as $\bs{a}$ gives $\bs{a}\perp\operatorname{span}(\bs{W})$ with $\bs{a}^{H}\bs{w}_{\rm f}\ne0$, and then $\rho(q\bs{a})=1+q^{2}P|\bs{a}^{H}\bs{w}_{\rm f}|^{2}/\sigma_{\rm w}^{2}\to\infty$ as $q\to\infty$.
Otherwise, $\bs{w}_{\rm f}=\bs{W}\bs{b}$ with the minimum-norm coefficient $\bs{b}=\bs{W}^{\dagger}\bs{w}_{\rm f}$.
If $\|\bs{b}\|^{2}>\varrho$, the channels $\bs{a}_{q}=q(\bs{W}^{H})^{\dagger}\bs{b}\in\operatorname{span}(\bs{W})\subseteq\mathcal{S}$ satisfy $\bs{W}^{H}\bs{a}_{q}=q\bs{b}$ and $\bs{a}_{q}^{H}\bs{w}_{\rm f}=q\|\bs{b}\|^{2}$, so $\rho(\bs{a}_{q})=1+q^{2}P\|\bs{b}\|^{4}/(q^{2}P\|\bs{b}\|^{2}+\sigma_{\rm w}^{2})\to1+\|\bs{b}\|^{2}>\rho^{\star}$.\hfill$\blacksquare$

For the passive deployment with all direct links blocked, $\sigma_{\rm w}^{2}$ does not depend on the channel, and every design can be taken in $\mathcal{S}=\operatorname{span}(\bs{G}^{H})$, the support of $\bs{a}_{\rm w}=\bs{G}^{H}\bs{\Theta}_{\rm T}^{H}\bs{h}_{\rm w}$ when every $\beta_i^{\rm T}>0$ and $\bs{R}_{h}$ is nonsingular, since a component in the null space of $\bs{G}$ reaches no node.
Part (\textit{ii}) thus applies, so that shielding is both necessary and sufficient for zero covert outage.

\emph{Proposition~\ref{prop:multi}.} Let $\bs{u}=\sqrt{P}\bs{\Sigma}_{0}^{-1/2}\bs{A}^{H}\bs{w}_{\rm f}$ and $\zeta=\|\bs{u}\|^{2}$, so that $\bs{\Sigma}_{0}^{-1/2}\bs{\Sigma}_{1}\bs{\Sigma}_{0}^{-1/2}=\bs{I}+\bs{u}\bs{u}^{H}$.
Let $\bs{U}$ be unitary with $\bs{U}^{H}\bs{u}=\sqrt{\zeta}\,\bs{e}_{1}$, where $\bs{e}_{1}$ is the first unit vector, and transform every received sample $\bs{y}_{i}$ as $\bs{z}_{i}=\bs{U}^{H}\bs{\Sigma}_{0}^{-1/2}\bs{y}_{i}$.
Under both hypotheses the entries $2,\ldots,N_{\rm w}$ of $\bs{z}_{i}$ are i.i.d. $\mathcal{CN}(0,1)$ and independent of the first entry, which is $\mathcal{CN}(0,1)$ under $\mathcal{H}_{0}$ and $\mathcal{CN}(0,1+\zeta)$ under $\mathcal{H}_{1}$.
The first entries of the $n$ samples are therefore a sufficient statistic, and the minimum DEP equals that of the scalar test between variances $1$ and $1+\zeta$ with $n$ samples, whose optimal detector is the radiometer, so $\mathrm{DEP}_{\min}=\mathrm{DEP}^{\star}(n,1+\zeta)$.
With $\bs{w}_{\rm f}=\bs{W}\bs{b}$ and $\bs{J}=\sqrt{P}\bs{A}^{H}\bs{W}$, $\zeta=\bs{b}^{H}\bs{J}^{H}(\bs{J}\bs{J}^{H}+\bs{\Sigma})^{-1}\bs{J}\bs{b}$, and $\bs{J}^{H}(\bs{J}\bs{J}^{H}+\bs{\Sigma})^{-1}\bs{J}=\bs{I}-(\bs{I}+\bs{J}^{H}\bs{\Sigma}^{-1}\bs{J})^{-1}\prec\bs{I}$, so $\zeta\le\|\bs{b}\|^{2}\le\varrho$, with equality only for $\bs{b}=\bs{0}$.
Since $\mathrm{DEP}^{\star}(n,\rho)$ decreases in $\rho$, $\mathrm{DEP}_{\min}\ge\mathrm{DEP}^{\star}(n,\rho^{\star})=1-\epsilon$.\hfill$\blacksquare$

\section{Proofs of Proposition~\ref{prop:outage} and Corollary~\ref{cor:hybrid}}\label{app:outage}
\emph{Proposition~\ref{prop:outage}.} $\bs{M}$ is a rank-one positive semidefinite matrix minus a positive semidefinite one, so by Weyl's inequality $\lambda_{2}(\bs{M})\le\lambda_{2}(\bs{q}_{\rm f}\bs{q}_{\rm f}^{H})=0$.
Since $\bs{R}_{h}$ is nonsingular, Sylvester's law of inertia gives the same inequality for the congruent matrix $\bs{K}=\bs{R}_{h}^{1/2}\bs{M}\bs{R}_{h}^{1/2}$, i.e., $\lambda_{2}(\bs{K})\le0$.
With $\bs{h}_{\rm w}=\bs{R}_{h}^{1/2}\bs{x}$, $\bs{x}\sim\mathcal{CN}(\bs{0},\bs{I})$ and $\bs{K}=\bs{U}\bs{\Lambda}\bs{U}^{H}$, the quadratic form equals $\sum_{j}\lambda_{j}E_{j}$ with $E_{j}=|(\bs{U}^{H}\bs{x})_{j}|^{2}$ i.i.d. unit exponentials.
If $\lambda_{1}\le0$, it never exceeds $\varrho\sigma_{\rm w}^{2}\ge0$.
Otherwise, conditioning on $E_{2},\ldots,E_{N}$, $\Pr\{\lambda_{1}E_{1}>\varrho\sigma_{\rm w}^{2}+\sum_{j\ge2}|\lambda_{j}|E_{j}\}=e^{-\varrho\sigma_{\rm w}^{2}/\lambda_{1}}\prod_{j\ge2}(1+|\lambda_{j}|/\lambda_{1})^{-1}$, which is \eqref{eq:pout}.\hfill$\blacksquare$

\emph{Corollary~\ref{cor:hybrid}.} (\textit{i}) $\bs{F}\bs{B}=\bs{\Pi}\bs{W}_{\rm all}$ with the projection $\bs{\Pi}=\bs{F}\bs{F}^{\dagger}$, so the realized covert precoder is $\bs{\Pi}\bs{w}_{\rm f}=(\bs{\Pi}\bs{W})\bs{b}$, which is \eqref{eq:shield} for the realized covers $\bs{\Pi}\bs{W}$.
A common scaling of the three columns leaves $\bs{b}$ unchanged.
(\textit{ii}) The first equality follows from $\bs{w}_{\rm f}=\bs{W}\bs{b}$.
Write $\bs{W}_{\rm all}=\bs{U}\bs{E}$, where the columns of $\bs{U}\in\Comp^{M\times L}$ form a basis of $\operatorname{span}(\bs{W})$ and $L=\operatorname{rank}\bs{W}$.
Every $\bs{u}\in\Comp^{M}$ is the sum of two constant-modulus vectors, $u_{m}=A_{0}e^{j(\phi_{m}+\psi_{m})}+A_{0}e^{j(\phi_{m}-\psi_{m})}$ with $\phi_{m}=\angle u_{m}$, $A_{0}=\max_{m}|u_{m}|/2$ and $\psi_{m}=\arccos(|u_{m}|/2A_{0})$ \cite{zhang2005variable, sohrabi2016hybrid}.
Hence $\bs{U}=\bs{F}_{0}\bs{D}$ with a constant-modulus $\bs{F}_{0}\in\Comp^{M\times2L}$ and a digital $\bs{D}$, and $\bs{W}_{\rm all}=\bs{F}_{0}(\bs{D}\bs{E})$.\hfill$\blacksquare$

\section{Proofs of Proposition~\ref{prop:rsma} and Corollary~\ref{cor:cf}, and Derivation of Remark~\ref{rem:crossover}}\label{app:rsma}
Write $X=Pp_{{\rm n},{\rm c}}$ and $Y=Pp_{{\rm n},{\rm n}}$ for the public user's received common and private powers.

\emph{Part (\textit{i}).} With $\bs{w}_{\rm n}=\bs{0}$, \eqref{eq:qos} requires $(1-\tau)R_{\rm c}\ge R_{\rm th}$, so $R_{\rm c}\ge R_{\rm th}$ and the same design with $\tau=0$ is feasible for (P1-N) at the same covert rate.
If $R_{\rm th}\le2r$, take a feasible design of (P1-S), set $\bs{w}_{\rm n}=\bs{0}$ and rescale $\bs{w}_{\rm c}$ by $s=(\|\bs{w}_{\rm c}\|^{2}(1+|c|^{2}))^{-1/2}\ge1$, keeping $c$.
The common SINR at each user does not decrease, so both users decode the common stream at a rate of at least $R_{\rm c}\ge2r\ge R_{\rm th}$.
The covert SINR $s^{2}|c|^{2}Pp_{{\rm f},{\rm c}}/\sigma^{2}$ is at least its former value, and the shield is unchanged.

\emph{Part (\textit{ii}).} With $p_{k,{\rm f}}=|c|^{2}p_{k,{\rm c}}$, $\gamma_{{\rm c},k}<1/|c|^{2}$.
In (P1-N), $R_{\rm c}\ge R_{\rm th}$ gives $2^{R_{\rm th}}-1<1/|c|^{2}$.
In (P1-S), let $\chi=(1-\tau)R_{\rm c}\ge r$ be the public user's common share.
The covert user's floor gives $R_{\rm c}\ge\chi+r$, and \eqref{eq:qos} gives $\log_{2}(1+\gamma_{\rm n})\ge[R_{\rm th}-\chi]^{+}$.
Since $\gamma_{\rm n}=Y/(|c|^{2}X+\sigma^{2})$, $Y\ge(2^{[R_{\rm th}-\chi]^{+}}-1)(|c|^{2}X+\sigma^{2})$, and decoding the common stream requires $X\ge(2^{\chi+r}-1)(Y+|c|^{2}X+\sigma^{2})>(2^{\chi+r}-1)2^{[R_{\rm th}-\chi]^{+}}|c|^{2}X$.
The factor is increasing in $\chi$, and for $R_{\rm th}\ge r$ its minimum over $\chi\ge r$ is $2^{R_{\rm th}-r}(2^{2r}-1)$.

\emph{Part (\textit{iii}).} $\bs{w}_{\rm f}=\bs{W}\bs{b}$ with $\bs{b}=b_{0}[1,-\nu]^{T}$ and $\|\bs{b}\|^{2}\le\varrho$, so Proposition~\ref{prop:shield}(\textit{i}) applies, and $\bs{a}_{\rm n}^{H}\bs{w}_{\rm f}=b_{0}(\bs{a}_{\rm n}^{H}\bs{w}_{\rm c}-\nu\bs{a}_{\rm n}^{H}\bs{w}_{\rm n})=0$.
Since $|\nu|^{2}=X/Y$ and the common stream must be decodable with the private stream as interference, $X/Y>X/(Y+\sigma^{2})\ge2^{R_{\rm c}}-1\ge2^{2r}-1$, hence $|b_{0}|^{2}\le\varrho /(1+X/Y)<\varrho 2^{-2r}$.\hfill$\blacksquare$

\emph{Remark~\ref{rem:crossover}.} In the high-SNR limit, the covert SINR is proportional to the admissible covert-to-cover ratio.
Under the assumption of the remark, each cap of Proposition~\ref{prop:rsma} is approached with the same cover power at the covert user.
Relative to the certificate level $\varrho$, NOMA loses $[\log_{2}(\varrho (2^{R_{\rm th}}-1))]^{+}$, all-common RSMA loses $[\log_{2}(\varrho (2^{R_{\rm th}+r}-1))]^{+}$, and the nulled shield loses $2r$, its cap being approached by $R_{\rm c}=2r$, $\tau=1/2$, and $X/Y\to2^{2r}-1$ for every $R_{\rm th}$.
Equating the losses gives \eqref{eq:crossover}.\hfill$\blacksquare$

\emph{Corollary~\ref{cor:cf}.} (\textit{i}) With $\bs{w}_{\rm n}=\bs{0}$, $p_{{\rm f},{\rm n}}=0$ and $\gamma_{\rm f}=P_{S}|c|^{2}|\bs{a}_{\rm f}^{H}\bs{w}_{\rm c}|^{2}/\sigma^{2}$, and the power normalization gives $\|\bs{w}_{\rm c}\|^{2}(1+|c|^{2})=1$.
By the Cauchy--Schwarz inequality, $|\bs{a}_{\rm f}^{H}\bs{w}_{\rm c}|^{2}\le\|\bs{a}_{\rm f}\|^{2}\|\bs{w}_{\rm c}\|^{2}=\|\bs{a}_{\rm f}\|^{2}/(1+|c|^{2})$, with equality if and only if $\bs{w}_{\rm c}\propto\bs{a}_{\rm f}$, which gives \eqref{eq:cfbound}.
Moreover, $|c|^{2}/(1+|c|^{2})$ increases in $|c|^{2}$, and $|c|^{2}\le\varrho$ by \eqref{eq:shield}.
(\textit{ii}) With $\bs{w}_{\rm n}=\bs{0}$, $R_{\rm n}=(1-\tau)R_{\rm c}$, so \eqref{eq:qos} requires $(1-\tau)R_{\rm c}\ge R_{\rm th}$ and $\tau R_{\rm c}\ge r$, hence $R_{\rm c}\ge R_{\rm th}+r$.
Since $p_{k,{\rm f}}=|c|^{2}p_{k,{\rm c}}$ and $p_{k,{\rm n}}=0$, $\gamma_{{\rm c},k}=Pp_{k,{\rm c}}/(P|c|^{2}p_{k,{\rm c}}+\sigma^{2})<1/|c|^{2}$, so $2^{R_{\rm th}+r}-1<1/|c|^{2}$.\hfill$\blacksquare$

\bibliographystyle{IEEEtran}
\bibliography{reference}

@article{bash2013limits,
  title={{Limits of Reliable Communication with Low Probability of Detection on AWGN Channels}},
  author={Bash, Boulat A. and Goeckel, Dennis and Towsley, Don},
  journal={IEEE J. Sel. Areas Commun.},
  volume={31},
  number={9},
  pages={1921--1930},
  year={2013},
  doi={10.1109/JSAC.2013.130923}
}

@article{yan2019gaussian,
  title={{Gaussian Signalling for Covert Communications}},
  author={Yan, Shihao and Cong, Yirui and Hanly, Stephen V. and Zhou, Xiangyun},
  journal={IEEE Trans. Wireless Commun.},
  volume={18},
  number={7},
  pages={3542--3553},
  year={2019},
  doi={10.1109/TWC.2019.2915305}
}

@article{vu2025performance,
  title={{On Performance of IoT Networks With Coordinated NOMA Transmission: Covert Monitoring and Information Decoding}},
  author={Vu, Thai-Hoc and Le, Anh-Tu and Tu, Ngo Hoang and Nguyen, Tan N. and Voznak, Miroslav},
  journal={IEEE Internet Things J.},
  volume={12},
  number={22},
  pages={48069--48084},
  year={2025},
  doi={10.1109/JIOT.2025.3605276}
}

@article{mu2022simultaneously,
  title={{Simultaneously Transmitting and Reflecting (STAR) RIS Aided Wireless Communications}},
  author={Mu, Xidong and Liu, Yuanwei and Guo, Li and Lin, Jiaru and Schober, Robert},
  journal={IEEE Trans. Wireless Commun.},
  volume={21},
  number={5},
  pages={3083--3098},
  year={2022},
  doi={10.1109/TWC.2021.3118225}
}

@article{mao2018rate,
  title={{Rate-splitting multiple access for downlink communication systems: bridging, generalizing, and outperforming SDMA and NOMA}},
  author={Mao, Yijie and Clerckx, Bruno and Li, Victor O. K.},
  journal={EURASIP J. Wireless Commun. Netw.},
  volume={2018},
  number={1},
  pages={133},
  year={2018},
  doi={10.1186/s13638-018-1104-7}
}

@inproceedings{tu2025active,
  title={{Active STAR-RIS-Empowered RSMA Systems: Outage Probability and Ergodic Capacity}},
  author={Tu, Ngo Hoang and Thanh, Tran Thien and Bao, Vo Nguyen Quoc},
  booktitle={Proc. 2025 RIVF Int. Conf. Comput. Commun. Technol. (RIVF)},
  pages={1239--1244},
  year={2025},
  doi={10.1109/RIVF68649.2025.11365149}
}

@article{xiao2024star,
  title={{STAR-RIS Enhanced Joint Physical Layer Security and Covert Communications for Multi-Antenna mmWave Systems}},
  author={Xiao, Han and Hu, Xiaoyan and Li, Ang and Wang, Wenjie and Su, Zhou and Wong, Kai-Kit and Yang, Kun},
  journal={IEEE Trans. Wireless Commun.},
  volume={23},
  number={8},
  pages={8805--8819},
  year={2024},
  doi={10.1109/TWC.2024.3354452}
}

@article{nguyen2025enhancing,
  title={{Enhancing Covertness in Hybrid Active-Passive STAR-RIS-Based RSMA Systems}},
  author={Nguyen, Lan Anh and Vu, Thai-Hoc and Kim, Sunghwan},
  journal={IEEE Trans. Veh. Technol.},
  volume={74},
  number={10},
  pages={16568--16573},
  year={2025},
  doi={10.1109/TVT.2025.3571459}
}

@article{chang2024star,
  title={{STAR-RIS-Aided Covert Communications in MISO-RSMA Systems}},
  author={Chang, Heng and Kang, Xueyu and Lei, Hongjiang and Tsiftsis, Theodoros A. and Pan, Gaofeng and Liu, Hongwu},
  journal={IEEE Trans. Green Commun. Netw.},
  volume={8},
  number={4},
  pages={1318--1331},
  year={2024},
  doi={10.1109/TGCN.2024.3432656}
}

@article{wang2026joint,
  title={{Joint Beamforming and Deployment Optimization for Active STAR-RIS Enabled Covert Communications}},
  author={Wang, Yawen and Fan, Jiancun and Xu, Hengbo and Yan, Qingze and Luo, Jie and Yan, Feng and Jia, Xiaolin},
  journal={IEEE Trans. Wireless Commun.},
  volume={25},
  pages={18471--18487},
  year={2026},
  doi={10.1109/TWC.2026.3694125}
}

@article{huang2020reconfigurable,
  title={{Reconfigurable Intelligent Surface Assisted Multiuser MISO Systems Exploiting Deep Reinforcement Learning}},
  author={Huang, Chongwen and Mo, Ronghong and Yuen, Chau},
  journal={IEEE J. Sel. Areas Commun.},
  volume={38},
  number={8},
  pages={1839--1850},
  year={2020},
  doi={10.1109/JSAC.2020.3000835}
}

@article{liang2025covert,
  title={{Covert Communications for Active STAR-RIS-Aided RSMA Systems With Hardware Impairments}},
  author={Liang, Yuting and Huang, Kewen and Yang, Liang and Li, Xingwang and Liu, Hongwu and Bian, Yougang},
  journal={IEEE Trans. Intell. Transp. Syst.},
  volume={26},
  number={7},
  pages={10036--10047},
  year={2025},
  doi={10.1109/TITS.2025.3542808}
}

@article{yu2026covert,
  title={{Covert Performance of STAR-RIS Aided THz Communication System With RSMA and Phase Errors}},
  author={Yu, Xiangbin and Zhou, Yue and Yan, Shihao and Rui, Yun and Dang, Xiaoyu and Yuen, Chau and Guizani, Mohsen},
  journal={IEEE J. Sel. Areas Commun.},
  volume={44},
  pages={1944--1959},
  year={2026},
  doi={10.1109/JSAC.2025.3638299}
}

@article{bai2023covert,
  title={{Covert Communication for Spatially Sparse mmWave Massive MIMO Channels}},
  author={Bai, Lin and Xu, Jinpeng and Zhou, Lin},
  journal={IEEE Trans. Commun.},
  volume={71},
  number={3},
  pages={1615--1630},
  year={2023},
  doi={10.1109/TCOMM.2023.3237998}
}

@inproceedings{ci2021hybrid,
  title={{Hybrid Beamforming Design for Covert Multicast mmWave Massive MIMO Communications}},
  author={Ci, Wei and Qi, Chenhao and Li, Geoffrey Ye and Mao, Shiwen},
  booktitle={Proc. IEEE GLOBECOM},
  pages={1--6},
  year={2021},
  doi={10.1109/GLOBECOM46510.2021.9685146}
}

@article{tu2025semi,
  title={{Semi-Static Hybrid Beamforming for O-RAN mmWave Massive MIMO Systems}},
  author={Tu, Ngo Hoang and Kim, Minhyun and Lee, Kyungchun},
  journal={IEEE Trans. Wireless Commun.},
  volume={24},
  number={11},
  pages={9791--9808},
  year={2025},
  doi={10.1109/TWC.2025.3575351}
}

@article{tu2026hybrid,
  title={{Hybrid Beamforming and Deep-Learning-Enabled Precoding for O-RAN mmWave Massive MIMO}},
  author={Tu, Ngo Hoang and Kim, Minhyun and Lee, Kyungchun},
  journal={IEEE Trans. Wireless Commun.},
  volume={25},
  pages={4053--4069},
  year={2026},
  doi={10.1109/TWC.2025.3607838}
}

@article{tu2026multi,
  title={{Multi-RIS-Aided Cell-Free mmWave Massive MIMO With Short-Packet uRLLC: Passive and Hybrid Active Beamforming}},
  author={Tu, Ngo Hoang and Lee, Kyungchun},
  journal={IEEE Trans. Wireless Commun.},
  volume={25},
  pages={14908--14924},
  year={2026},
  doi={10.1109/TWC.2026.3680903}
}

@article{sohrabi2016hybrid,
  title={{Hybrid Digital and Analog Beamforming Design for Large-Scale Antenna Arrays}},
  author={Sohrabi, Foad and Yu, Wei},
  journal={IEEE J. Sel. Topics Signal Process.},
  volume={10},
  number={3},
  pages={501--513},
  year={2016},
  doi={10.1109/JSTSP.2016.2520912}
}

@article{tao2020covert,
  title={{Covert Communication in Downlink NOMA Systems With Random Transmit Power}},
  author={Tao, Liwei and Yang, Weiwei and Yan, Shihao and Wu, Dan and Guan, Xinrong and Chen, Dechuan},
  journal={IEEE Wireless Commun. Lett.},
  volume={9},
  number={11},
  pages={2000--2004},
  year={2020},
  doi={10.1109/LWC.2020.3011191}
}

@article{le2025active,
  title={{Active-Reconfigurable-Repeater-Assisted NOMA Networks in Internet of Things: Reliability, Security, and Covertness}},
  author={Le, Anh-Tu and Vu, Thai-Hoc and Tu, Ngo Hoang and Nguyen, Tan N. and Tu, Lam-Thanh and Voznak, Miroslav},
  journal={IEEE Internet Things J.},
  volume={12},
  number={7},
  pages={8759--8772},
  year={2025},
  doi={10.1109/JIOT.2024.3503278}
}

@article{tu2026active,
  title={{Active STAR-RIS-Aided RSMA IoT Systems: Performance Analysis, Model-Based and Data-Driven Resource Allocation Frameworks}},
  author={Tu, Ngo Hoang and Bao, Vo Nguyen Quoc and Thanh, Tran Thien},
  journal={IEEE Internet Things J.},
  year={2026},
  note={early access, doi: 10.1109/JIOT.2026.3748576},
  doi={10.1109/JIOT.2026.3748576}
}

@article{wu2021intelligent,
  title={{Intelligent Reflecting Surface (IRS)-Aided Covert Communication With Warden's Statistical CSI}},
  author={Wu, Chaoyi and Yan, Shihao and Zhou, Xiaobo and Chen, Riqing and Sun, Jiande},
  journal={IEEE Wireless Commun. Lett.},
  volume={10},
  number={7},
  pages={1449--1453},
  year={2021},
  doi={10.1109/LWC.2021.3069778}
}

@article{yang2026intelligent,
  title={{Intelligent Covert ISAC via RIS: A Reinforcement Learning Approach}},
  author={Yang, Fangtao and Xing, Chengwen and Wei, Haichao and Jo, Minho and Deng, Na and Zhao, Nan and Niyato, Dusit},
  journal={IEEE Trans. Wireless Commun.},
  volume={25},
  pages={13107--13120},
  year={2026},
  doi={10.1109/TWC.2026.3667559}
}

@article{hieu2023joint,
  title={{Joint Power Allocation and Rate Control for Rate Splitting Multiple Access Networks With Covert Communications}},
  author={Hieu, Nguyen Quang and Hoang, Dinh Thai and Niyato, Dusit and Nguyen, Diep N. and Kim, Dong In and Jamalipour, Abbas},
  journal={IEEE Trans. Commun.},
  volume={71},
  number={4},
  pages={2274--2287},
  year={2023},
  doi={10.1109/TCOMM.2023.3242670}
}

@article{amiri2025resource,
  title={{Resource Allocation in STAR-RIS-Aided SWIPT With RSMA via Meta-Learning}},
  author={Amiri, Mojtaba and Vaezpour, Elaheh and Javadi, Sepideh and {Robat Mili}, Mohammad and Bennis, Mehdi and Jorswieck, Eduard A.},
  journal={IEEE Open J. Commun. Soc.},
  volume={6},
  pages={3806--3815},
  year={2025},
  doi={10.1109/OJCOMS.2025.3556484}
}

@article{wang2022intelligent,
  title={{Intelligent Reflecting Surface Configurations for Smart Radio Using Deep Reinforcement Learning}},
  author={Wang, Wei and Zhang, Wei},
  journal={IEEE J. Sel. Areas Commun.},
  volume={40},
  number={8},
  pages={2335--2346},
  year={2022},
  doi={10.1109/JSAC.2022.3180787}
}

@article{li2026star,
  title={{STAR-RIS-Assisted Covert Communications in RSMA Networks: A Quantum Reinforcement Learning Approach}},
  author={Li, Jiameng and Wang, Xiaojie and Wu, Jun and Ning, Zhaolong and Guo, Song},
  journal={IEEE Trans. Mobile Comput.},
  volume={25},
  number={8},
  pages={11723--11740},
  year={2026},
  doi={10.1109/TMC.2026.3666828}
}

@article{shiu2000fading,
  title={{Fading correlation and its effect on the capacity of multielement antenna systems}},
  author={Shiu, Da-Shan and Foschini, Gerard J. and Gans, Michael J. and Kahn, Joseph M.},
  journal={IEEE Trans. Commun.},
  volume={48},
  number={3},
  pages={502--513},
  year={2000},
  doi={10.1109/26.837052}
}

@inproceedings{silver2014deterministic,
  title={{Deterministic Policy Gradient Algorithms}},
  author={Silver, David and Lever, Guy and Heess, Nicolas and Degris, Thomas and Wierstra, Daan and Riedmiller, Martin},
  booktitle={Proc. Int. Conf. Mach. Learn. (ICML)},
  volume={32},
  pages={387--395},
  year={2014},
  series={PMLR}
}

@inproceedings{fujimoto2018addressing,
  title={{Addressing Function Approximation Error in Actor-Critic Methods}},
  author={Fujimoto, Scott and van Hoof, Herke and Meger, David},
  booktitle={Proc. Int. Conf. Mach. Learn. (ICML)},
  volume={80},
  pages={1587--1596},
  year={2018},
  doi={10.48550/arXiv.1802.09477},
  series={PMLR}
}

@inproceedings{haarnoja2018soft,
  title={{Soft Actor-Critic: Off-Policy Maximum Entropy Deep Reinforcement Learning with a Stochastic Actor}},
  author={Haarnoja, Tuomas and Zhou, Aurick and Abbeel, Pieter and Levine, Sergey},
  booktitle={Proc. Int. Conf. Mach. Learn. (ICML)},
  volume={80},
  pages={1861--1870},
  year={2018},
  doi={10.48550/arXiv.1801.01290},
  series={PMLR}
}

@misc{schulman2017proximal,
  title={{Proximal Policy Optimization Algorithms}},
  author={Schulman, John and Wolski, Filip and Dhariwal, Prafulla and Radford, Alec and Klimov, Oleg},
  year={2017},
  note={arXiv:1707.06347},
  eprint={1707.06347},
  archiveprefix={arXiv},
  primaryclass={cs.LG}
}

@article{mnih2015human,
  title={{Human-Level Control through Deep Reinforcement Learning}},
  author={Mnih, Volodymyr and Kavukcuoglu, Koray and Silver, David and Rusu, Andrei A. and Veness, Joel and Bellemare, Marc G. and Graves, Alex and Riedmiller, Martin and others},
  journal={Nature},
  volume={518},
  number={7540},
  pages={529--533},
  year={2015},
  doi={10.1038/nature14236}
}

@misc{ahmadi2026square,
  title={Square-Root Law for Covert Communication with Warden-Favorable Side Information},
  author={Hossein Ahmadi and Christian Deppe and Boulat A. Bash and Eduard A. Jorswieck},
  year={2026},
  note={arXiv:2607.14013}
}

@article{zhang2023active,
  title={{Active RIS vs. Passive RIS: Which Will Prevail in 6G?}},
  author={Zhang, Zijian and Dai, Linglong and Chen, Xibi and Liu, Changhao and Yang, Fan and Schober, Robert and Poor, H. Vincent},
  journal={IEEE Trans. Commun.},
  volume={71},
  number={3},
  pages={1707--1725},
  year={2023},
  doi={10.1109/TCOMM.2022.3231893}
}

@article{zhang2005variable,
  title={{Variable-phase-shift-based RF-baseband codesign for MIMO antenna selection}},
  author={Zhang, Xinying and Molisch, Andreas F. and Kung, Sun-Yuan},
  journal={IEEE Trans. Signal Process.},
  volume={53},
  number={11},
  pages={4091--4103},
  year={2005},
  doi={10.1109/TSP.2005.857024}
}

@article{xia2020deep,
  title={{A Deep Learning Framework for Optimization of MISO Downlink Beamforming}},
  author={Xia, Wenchao and Zheng, Gan and Zhu, Yongxu and Zhang, Jun and Wang, Jiangzhou and Petropulu, Athina P.},
  journal={IEEE Trans. Commun.},
  volume={68},
  number={3},
  pages={1866--1880},
  year={2020},
  doi={10.1109/TCOMM.2019.2960361}
}

@article{gao2020unsupervised,
  title={{Unsupervised Learning for Passive Beamforming}},
  author={Gao, Jiabao and Zhong, Caijun and Chen, Xiaoming and Lin, Hai and Zhang, Zhaoyang},
  journal={IEEE Commun. Lett.},
  volume={24},
  number={5},
  pages={1052--1056},
  year={2020},
  doi={10.1109/LCOMM.2020.2965532}
}

@book{sutton2018reinforcement,
  title={Reinforcement Learning: An Introduction},
  author={Sutton, Richard S. and Barto, Andrew G.},
  edition={2nd},
  year={2018},
  publisher={MIT Press},
  address={Cambridge, MA, USA}
}

@inproceedings{heess2015learning,
  title={Learning continuous control policies by stochastic value gradients},
  author={Heess, Nicolas and Wayne, Greg and Silver, David and Lillicrap, Timothy and Erez, Tom and Tassa, Yuval},
  booktitle={Proc. Adv. Neural Inf. Process. Syst. (NeurIPS)},
  year={2015}
}

@book{rappaport2002wireless,
  author    = {Theodore S. Rappaport},
  title     = {Wireless Communications: Principles and Practice},
  edition   = {2nd},
  publisher = {Prentice Hall},
  address   = {Upper Saddle River, NJ, USA},
  year      = {2002}
}

@IEEEtranBSTCTL{IEEEexample:BSTcontrol,
  CTLuse_forced_etal = "yes",
  CTLmax_names_forced_etal = "6",
  CTLnames_show_etal = "1"
}

\end{document}